\documentclass[astrosymb,trackchanges]{aastex7} %
\usepackage{physics,amsmath} %
\usepackage{bm} %  boldsymbol
\renewcommand{\vec}{\boldsymbol} %
\newcommand{\HI}{\mbox{H\thinspace{\sc i} }} %
\newcommand{\HII}{\mbox{H\thinspace{\sc ii} }} %
\newcommand{\dif}{\mathrm{d}} %
\shorttitle{Pinched Magnetic Fields in W3~IRS5} %
\shortauthors{Chen et al.} %
\graphicspath{{./}{figures/}} %
\begin{document} %

\title{Pinched Magnetic Fields in the High-mass Protocluster W3~IRS5} %

\correspondingauthor{Huei-Ru Vivien Chen} %
%\email{hchen@phys.nthu.edu.tw} %

\author[orcid=0000-0002-9774-1846,sname='Chen']{Huei-Ru Vivien Chen} %
\affiliation{Institute of Astronomy and Department of Physics, National Tsing Hua University, Hsinchu 300044, Taiwan} %below authorship
\email[show]{hchen@phys.nthu.edu.tw}  

\author[orcid=0000-0003-2384-6589,sname='Zhang']{Qizhou Zhang} %
\affiliation{Center for Astrophysics $\vert$ Harvard \& Smithsonian, 60 Garden Street, Cambridge, MA, 02138, USA} % below authorship
\email{qzhang@cfa.harvard.edu} % hidden email 

\author[orcid=0000-0001-8516-2532,sname='Ching']{Tao-Chung Ching} %
\altaffiliation{Jansky Fellow of the National Radio Astronomy Observatory} % in footnote
\affiliation{National Radio Astronomy Observatory, 1003 Lopezville Road, Socorro, NM 87801, USA} % below authorship
%\email[show]{} % shown email
\email{ chingtaochung@gmail.com} % hidden email 

\author[orcid=0000-0002-1700-090X,sname='Beuthor']{H. Beuther} %
%\altaffiliation{} % in footnote
\affiliation{Max Planck Institute for Astronomy, K\"{o}nigstuhl 17, 69117 Heidelberg, Germany} % below authorshipi
%\email[show]{} % shown email
\email{beuther@mpia.de} % hidden email 

\author[orcid=0000-0002-9323-0950,sname='Wang']{Kuo-Song Wang} %
\affiliation{Institute of Astronomy and Astrophysics, Academia Sinica, Taipei 106319, Taiwan} % below authorship
\email{kswang@asiaa.sinica.edu.tw} % hidden email 

%\author[orcid=,sname='']{} %
%\altaffiliation{} % in footnote
%\affiliation{} % below authorship
%\email[show]{} % shown email
%\email{} % hidden email 

%\collaboration{all}{The Terra Mater collaboration}

\begin{abstract} %
We present polarization maps of dust emission at 340~GHz in the luminous high-mass protocluster, W3~IRS5, observed with the Submillimeter Array.
%with an angular resolution of $\sim 2\farcs4$.   
The projected magnetic fields appear fairly organized with a pinched morphology in the northern part and a concave shape in the southern part.  
We fit the polarization maps with a two-component magnetic field model: an hourglass model centered at the continuum peak, SMM2, and an empirical sphere centered at the O-type star, IRS7.  
Using the Davis-Chandrasekhar-Fermi method, we calculate a projected field strength of $B_\mathrm{pos} = 1.4 \; \mathrm{mG}$.   
Along with the Zeeman measurement, a total magnetic field strength of $B_\mathrm{tot} = 1.6 \; \mathrm{mG}$ is obtained.  
We find that the gravitational energy is the most dominant, followed by magnetic energy, and then turbulent energy. 
Small values of the virial parameter, $\alpha_\mathrm{vir} = 0.8$, and the ratio of timescales, $t_\mathrm{ff}/t_\mathrm{corss} = 0.6$, suggest an ongoing collapse.  
We also show collimated molecular outflows in the $\mathrm{CO \; (3-2)}$ and $\mathrm{SiO \; (8-7)}$ transitions.  
The morphology of magnetic fields and the surrounding \HII regions put forward a scenario for W3~IRS5.
A gravitationally unstable dense core formed within a neutral gas ridge plowed by the expansions of W3~A and W3~B.  
The core began to contract, causing gravity to pull the magnetic field lines inward, which resulted in a pinched field morphology.  
Subsequent expansion of W3~F, ionized by IRS7, perturbed the magnetic field, creating concave patterns.   
The dynamical interactions among protostars led to misalignment of their outflows. 
%To assess the relative importance of gravity, turbulence, and magnetic fields, we compare their energy contributions and find that gravitational energy is the most dominant, followed by magnetic energy, and then turbulent energy.
%Additionally, two tentative velocity gradients in distinct directions near SMM2 are discussed.  
\end{abstract} %



\section{Introduction} \label{sec:intro} %
Magnetic fields, along with turbulence, provide support to oppose gravitational contraction in molecular clouds and influence the dynamics and timescales of star formation \citep[e.g.][]{Shu1987,McKee2007}.
%The support of magnetic fields is not trivial but acts merely on ions through the Lorentz force and slowly propagates to neutrals via collisions. 
%The different responses of ions and neutrals to the magnetic field result in a process known as ambipolar diffusion \citep{Mestel1956}, where neutrals contract relative to ions that remain tied to magnetic field lines.  
%Even when the magnetic field threading the clump is strong enough to prevent an immediate collapse, ambipolar diffusion can eventually lead to the formation of a dense core that is gravitationally unstable. 
In the standard paradigm of isolated star formation \citep{Shu1987}, magnetically supported molecular clouds are expected to slowly contract through ambipolar diffusion \citep{Mestel1956}, wherein neutrals contract relative to ions that remain tied to magnetic field lines, and eventually form a dense core that is gravitationally unstable.
%In a cluster environment with high-mass stars, an expanding \HII region may drive ionization and shock fronts into adjacent molecular clouds and sweep up on its edge a layer of dense neutral postshock material that eventually becomes gravitationally unstable \citep{Elmegreen1977}.   
%Such shock fronts can also compress pre-existing condensations, making them gravitationally unstable.   
Alternatively, dense cores may form in a layer of sweep-up neutral material on the edge of expanding \HII regions when high-mass stars drive ionization and shock fronts into adjacent molecular clouds \citep{Elmegreen1977}. 
Once a dense core starts to collapse, the magnetic field lines are pulled inward under the flux freezing assumption, resulting in a pinched shape often described as an hourglass \citep{Li1996}.  
The core may fragment further as the central density increases and forms a small star group, in which low-mass stars reside, and perhaps intermediate-mass stars, and occasionally OB stars.  
Subsequent dynamical interactions among protostars may cause the individual disk-outflow systems to deviate from their initial configurations \citep{Zhang2014}.  

Linear polarized emission arising from magnetically aligned dust grains is currently the best available tool to map the morphology of the projected magnetic fields on the plane of sky, $B_\mathrm{pos}$ \citep{Pattle2023}.  
In the presence of a magnetic field, paramagnetic grains of elongated shape tend to spin with their minor axis aligned with the magnetic field direction, resulting in polarized emission in the millimeter wavebands perpendicular to the magnetic field \citep{Lazarian2007b,Hoang2008,Hoang2009,Andersson2015}. 
Polarization observations of high-mass star-forming regions suggest that magnetic fields are dynamically important during the formation of dense cores at spatial scales of $0.01 - 0.1 \; \mathrm{pc}$ \citep{Zhang2014}.  
The pinched morphology has been reported in many high-mass star-forming cores \citep[e.g.][]{Girart2009,Qiu2014,Beltran2019,Sanhueza2021,Saha2024,Beltran2024}.  
Unlike their low-mass counterparts, these dense cores often harbor multiple protostars, sometimes associated with hypercompact (HC) \HII regions.  
Stellar feedback, such as outflows and ionization of \HII regions, may further perturb or reshape the magnetic fields  \citep{Zapata2024}. 
It is unclear whether magnetic fields play an important role and how long the fields remain significant in the formation of stars across the entire range of masses.  

Located in the Perseus arm at a distance of $1.83 \pm 0.14 \; \mathrm{kpc}$ \citep{Imai2000}, the luminous \citep[$2 \times 10^5 \; L_\odot$, equivalent to an O6.5 star;][]{Campbell1995} infrared source, W3~IRS5, is the most massive star-forming core \citep[$715 \; M_\odot$;][]{Wang2012} in the W3~Main high-mass star-forming region.    
Its deeply embedded nature has attracted observations with the best angular resolutions available in the near-infrared (NIR) and centimeter wavebands \citep[Fig.~\ref{fig:bvec}a; e.g.][]{Imai2000,Wilson2003,Megeath2005,vanderTak2005c,Wang2013}. 
A small cluster of protostars, including two hypercompact (HC) \HII regions, has been identified within a projected separation of $5.6 \times 10^3 \; \mathrm{au}$ \citep{Wilson2003,Megeath2005,vanderTak2005c}.  
The inferred stellar surface density of $10^4\; \mathrm{pc^{-2}}$ is exceptionally high, so that \citet{Megeath2005} proposed W3~IRS5 to be a Trapezium cluster in the making.  
Using the Submillimeter Array (SMA), \citet{Wang2013} resolved dust continuum emission into five compact sources (blue cross, Fig.~\ref{fig:bvec}a), among which two (SMM1 and SMM2) coincide with the known HC \HII regions and three are newly identified with different chemical properties.  
Multiple outflows nearly perpendicular to each other have been reported: a large-scale CO bipolar outflow \citep[$\mathrm{P. A.} = 40^\circ$;][]{Mitchell1991} and three collimated bipolar outflows, including one lying close to the line of sight \citep{Wang2012,Rodon2008}.  
Proper motions of water masers and radio compact knots in $7 \; \mathrm{mm}$ continuum emission revealed high-speed jet flows in two  molecular outflows \citep{Imai2000,Wilson2003}.  
Figure~\ref{fig:bvec}a shows the axes of known outflows and their associated tracers.  
In addition to the embedded HC \HII regions, W3~IRS5 is surrounded by a cluster of \HII regions within merely $10^{\prime\prime}$ separation: the W3~A compact (C) \HII region to the east, the W3~B C\HII region to the west, and the W3~F ultracompact (UC) \HII region, ionized by the infrared source IRS~7, to the south \citep[blue star;][]{Wynn-Williams1972,Tieftrunk1997,Bik2012}.  
%From the $H66\alpha$ recombination line data, \citep{Tieftrunk1997} found a velocity difference of $\approx 5 \; \mathrm{km \, s^{-1}}$ across W3~F and suggested a scenario of a bow shock created by a star moving toward the dense molecular core surrounding W3~IRS5.  
Based on this rich population of \HII regions at different evolutionary stages, \citet{Tieftrunk1997} propose that massive star formation in W3~Main may be sequentially triggered by the pressure of the expanding \HII regions.   
Additionally, a complementary study with NIR data by \citet{Bik2012} found an age spread of $2-3 \; \mathrm{Myr}$ in the stellar content of W3~Main, which supports the scenario of sequential star formation.  

In the (sub)millimeter wavebands, dust polarized emission in W3~Main has been mapped at clump scales $\sim 0.2 \mathrm{pc}$ ($\sim 20^{\prime\prime}$) with single-dish telescopes \citep{Matthews2009,Dotson2010}. 
The inferred magnetic fields show a general orientation along $\mathrm{P.A.} = 140^\circ$ in the vicinity of W3~IRS5 (Fig.~\ref{fig:bvec}b) but start turning to east-west orientations in the W3~B compact \HII region \citep[see][]{Hull2014}.
The overall magnetic field pattern does not concur with a single large-scale ($\sim 1^\prime$) hourglass model at $\mathrm{P.A. \sim 50^\circ}$ proposed in earlier studies \citep{Roberts1993,Greaves1994}.  
Subsequent interferometric studies from the TADPOL/CARMA survey \citep{Hull2014} at $1.3 \; \mathrm{mm}$ greatly improved angular resolution to $2.9^{\prime\prime}$ ($0.026 \; \mathrm{pc}$) and revealed partly organized magnetic field patterns in W3~IRS5: a circular shape in the southern region and perhaps a partial hourglass shape in the northern region.  
\citet{Houde2016} applied the angular dispersion analysis, a technique not needing an assumed magnetic field geometry, to the TADPOL data and estimated a magnetic field strength of $B_\mathrm{pos} \sim 0.7 \; \mathrm{mG}$ in W3~IRS5.  
With all the extensive efforts to unveil its nature, W3~IRS5 remains enigmatic, and polarimetric observations with better angular resolutions may offer more insights into its magnetic fields. 
It should also be noted that W3~IRS5 is located high in the northern sky and will never be accessible with ALMA (Acatcama Large Millimeter/submillimeter Array).

We present SMA polarimetric observations of W3~IRS5 at $340 \; \mathrm{GHz}$ with an angular resolution of $2\farcs7 \times 2\farcs2$ that shows a pinched magnetic field morphology.  
This paper is organized as follows: in Sect~\ref{sec:obsdata}, we describe the SMA observations and data reduction; in Sect~\ref{sec:result}, we present the polarized dust emission, molecular outflows, and tentative detection of two velocity gradients; in Sect~\ref{sec:analysis}, we model the magnetic field and estimate its strength; in Sect.~\ref{sec:discussion}, we investigate the properties of the magnetic field and its importance relative to gravity and turbulence, then propose a formation scenario, and summarize the conclusions in Sect.~\ref{sec:summary}.  
Additionally, the observed and synthetic Stokes-$I$, $Q$, and $U$ maps are present in Appendix~\ref{sec:iqumap}.  
Tentative velocity gradients found with the SMA archival data are present in Appendix~\ref{sec:so2}. 
The radiative transfer equations used to calculate the Stokes $I$, $Q$, and $U$ images are explained in Appendix~\ref{sec:iqueqn}. 
The calculation of the virial mass for a power-law density distribution with a central star is in Appendix~\ref{sec:mvir}.

\iffalse %
\citet{Wilson2003} d1/d2 (SMM1) and b(SMM2) B0.5-B1, equivalent to $M_* \lesssim 12 M_\sun$. 
IRS~N7 is YSO not associated with a detectable \HII region or submiilimeter counter part \citep{Bik2012}. 
W3~F is possibly ionized by the infrared source IRS~7, a spectral type B0--O9.5 ZAMS star \citep{Wynn-Williams1972,Tieftrunk1997,Bik2012}. 
From the $H66\alpha$ recombination line data, \citep{Tieftrunk1997} found a velocity difference of $\approx 5 \; \mathrm{km \, s^{-1}}$ across the \HII region and suggested a scenario of a bow shock created by a star moving toward the dense molecular core surrounding W3~IRS5.  
The star powering W3~F may carve a photoionized arc into the denser neutral gas surrounding W3~IRS~5, causing the concave-shape distortion of the magnetic fields.  
The bright limb in the northern edge may indicate an ionization front created by the ram pressure of IRS~7.  
The thermal pressure of the high electron gas temperature at $T_e \approx 10^4 \; \mathrm{K}$ may produce a champagne flow toward the lower density gas in the south.  
Systemic velocity $v_\mathrm{LSR} = -38.2 \; \mathrm{km s^{-1}}$ \citep{Wang2013}.
Large-scale CO outflow reported by \citet{Mitchell1991} approximately at P. A. $ = 38^\circ$
\fi %

%%%%%%%%%%%%%
\section{Observations and Data Reduction \label{sec:obsdata}} %
Polarimetric observations with the SMA \citep{Ho2004} were carried out in the $345 \; \mathrm{GHz}$ band on 2012 August 7 and August 9 in the subcompact configuration with six antennas and 2012 November 7 in the compact configuration with seven antennas as part of the SMA Legacy Survey of \citet{Zhang2014}.  
The observations were performed with $4\;\mathrm{GHz}$ bandwidth in each sideband, separated by an IF frequency of $4-8 \; \mathrm{GHz}$.  
The phase center is at $(\alpha,\delta)(J2000) =$ (2:25:40.770,62:5:52.496).
The correlators were configured to have a uniform resolution of 128 channels per 104~MHz chunk, providing a channel width of $0.8 \; \mathrm{MHz}$ or $0.65 \; \mathrm{km \, s^{-1}}$.  
The receivers were tuned to the $\mathrm{CO} \; (3-2)$ line in the upper sideband.  
The combined visibility data cover a range of spatial frequencies from $7.5 \; \mathrm{k}\lambda$ to $76 \; \mathrm{k}\lambda$, which is insensitive to structures larger than $12^{\prime\prime}$, equivalent to $0.11 \; \mathrm{pc}$.  

The raw visibilities were inspected and calibrated for flux and bandpass in the IDL superset MIR, and were exported to MIRIAD for polarization calibration and imaging.  
The instrumental polarization response, i.e. the ``leakage'' term, for each antenna was calibrated by observing strong quasars, 3C84 and BL~Lac, over their transits to cover a wide hour angle range for a good parallactic angle coverage.  
The bandpass calibration was performed with BL~Lac.  
Uranus was used to set the flux density scale, which is estimated to be accurate within 15\%.  
Visibility data in both sidebands were used to generate line-free continuum maps in a multi-frequency synthesis, making an effective central frequency of $339.964 \; \mathrm{GHz}$ and a beam size of $2\farcs7 \times 2\farcs2$ ($\mathrm{P.A.} = 3.0^\circ$) with natural weighting. 
The resultant maps have an rms noise level of $\sigma_I = 21\; \mathrm{mJy \, beam^{-1}}$ for the Stokes~$I$ map and  $\sigma_P = 1.7\; \mathrm{mJy \, beam^{-1}}$ for maps of the Stokes~$Q$ and $U$ .  
The dynamic range of the Stokes~$I$ map is around $63 \, \sigma_I$.  
Once Stokes~$Q$, and $U$ are known, the polarization position angle (P.A.), $\psi$ can be deduced.  
The raw polarization intensity, $P_\mathrm{raw} = (Q^2 + U^2)^{1/2}$, is always positive and biased by the rms noise of Stokes~$Q$ and $U$, $\sigma_P$.
Instead, a debiased polarized intensity, $P = (Q^2 + U^2 - \sigma_P^2)^{1/2}$, is used \citep{Vaillancourt2006} when computing the fractional linear polarization, $P/I$, and throughout the study. 
%when computing the polarization percentage, $p = I_p/I$.  
The polarization position angle (P.A.) is determined by $\psi_\mathrm{obs} = (1/2) \arctan(U/Q)$, with an uncertainty of $\sigma_{\psi_\mathrm{obs}} = (1/2) (\sigma_P/P)$. 
After generating maps of Stokes $I$, $Q$, and $U$ (Fig.~\ref{fig:iqu}), we image the distribution of polarization, including $P$ and $\psi_\mathrm{obs}$, for regions where $P > 2 \sigma_P$ and $I > 3 \sigma_I$ (Fig.~\ref{fig:bvec}b).  
The threshold at $2$--$3\sigma$ is often applied on polarization maps to exclude measurements with large uncertainties in position angle and fractional linear polarization \citep[e.g.][]{Hull2014,Zhang2014,Beuther2020a}. 

In addition to the $\mathrm{CO} \; (3-2)$ transition, the passbands also cover many molecular lines, such as $\mathrm{SiO \; (8-7)}$, $\mathrm{H^{13}CN} \; (4-3)$, and several lines of $\mathrm{SO}$, $\mathrm{^{34}SO}$, $\mathrm{SO_2}$, and $\mathrm{^{34}SO_2}$.  
The spectral data were gridded with a velocity resolution of $0.8 \; \mathrm{km \, s^{-1}}$.  
Imaging with natural weighting yields a typical beam size of $2\farcs7 \times 2\farcs2$ and an rms noise level of $0.11 \; \mathrm{Jy \, beam^{-1}}$.  
For molecular lines used to trace core kinematics, such as $\mathrm{H^{13}CN} \; (4-3)$, we generate images with only visibilities from the compact configuration to alleviate the distortion caused by the lack of short spacing data.
Imaging with natural weighting produces a beam size of $2\farcs3 \times 1\farcs9$ and an rms noise level of $0.13 \; \mathrm{Jy \, beam^{-1}}$. 

%%%%%%%%%%%%%
\section{Observational Results \label{sec:result}} %
The 340~GHz continuum map is present in Fig.~\ref{fig:bvec}a along with the VLA~6cm continuum map showing the nearby \HII regions, W3~A, W3~B, and W3~F.     
The continuum emission (Fig.~\ref{fig:bvec}a) shows a single peak with extended emission similar to the morphology of the 223~GHz continuum map from the TADPOL survey \citep{Hull2014}.   
The extended emission to the west appears to truncate at the boundary of W3~B \citep[VLA $6 \; \mathrm{cm}$ map;][]{Tieftrunk1997}, possibly due to the lack of short spacing data.  
The five compact submillimeter sources \citep[blue crosses;][]{Wang2013} are located within the central peak.  
Multiple outflow axes associated with various tracers in the literature \citep{Mitchell1991,Imai2000,Wilson2003,Rodon2008,Wang2012} are also shown.  
The total flux density at 340~GHz is $7.1 \; \mathrm{Jy}$, which is 18\% of the total flux in the SCUBA $850 \; \mu\mathrm{m}$ map \citep{Wang2012}. 

\subsection{Pinched Magnetic Field Morphology} %
\begin{figure}[h!] %
\epsscale{0.6} %
\plotone{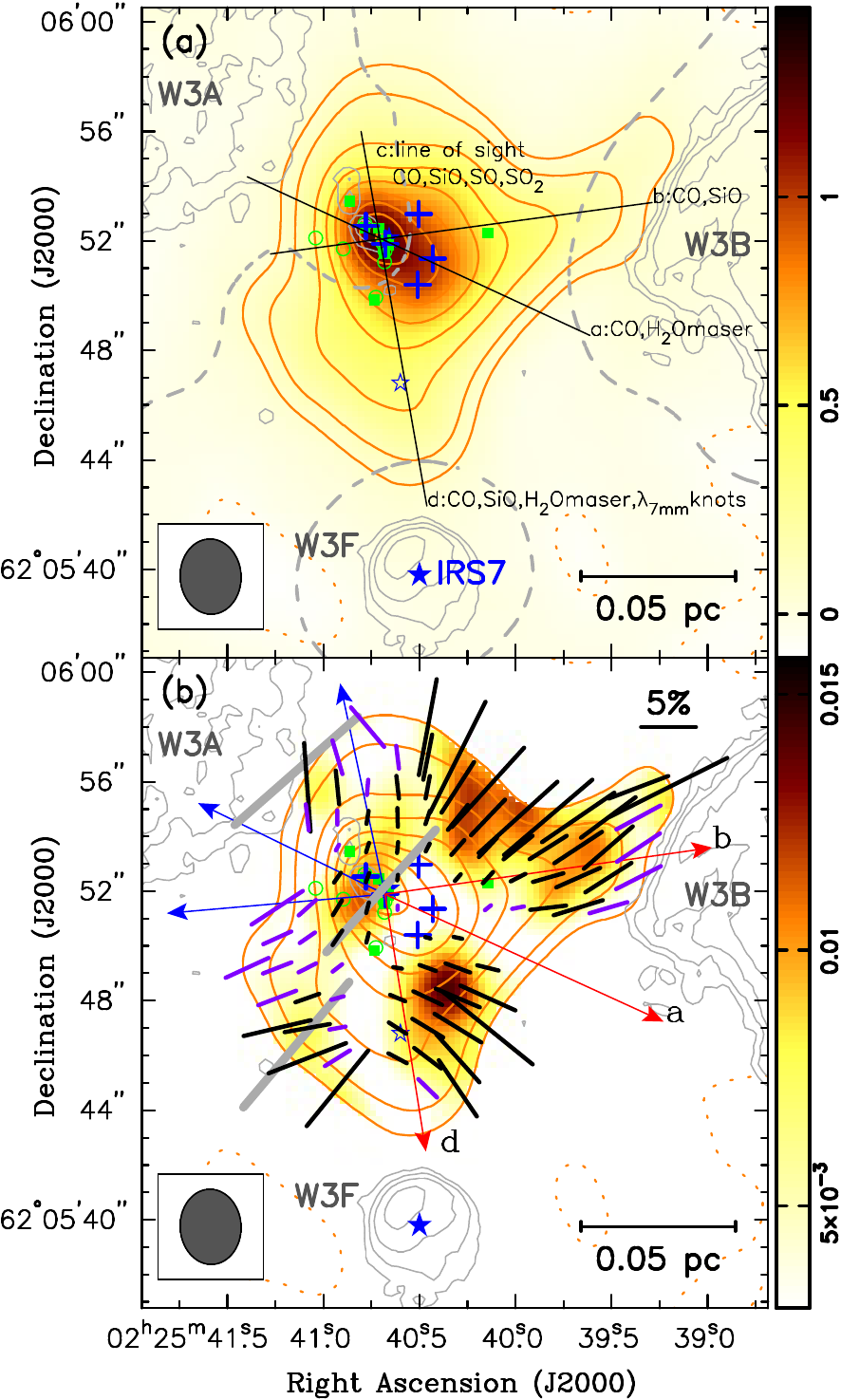} %
\caption{(a) SMA 340~GHz continuum map (color scale and red contours) of W3~IRS5 overlaid on VLA $6 \; \mathrm{cm}$ continuum map \citep[gray contours; ][]{Tieftrunk1997} with labels of the nearby \HII regions, W3~A, W3~B, and W3~F.  
Red contours are plotted at $(-3,3,6,14,22,30,38,46,54,62) \times \sigma$, where $\sigma = 21 \; \mathrm{mJy \, beam^{-1}}$ with a beam size of $2\farcs7 \times 2\farcs2$  ($\mathrm{P. A.} = 3^\circ$).    
Blue crosses mark the positions of five compact continuum sources identified at $354 \; \mathrm{GHz}$ \citep{Wang2013}, where SMM2 coincides with the continuum peak, and SMM1 is located to the east.  
Green open circles are the positions of near-infrared sources \citep{Megeath2005}, and green filled squares are the positions of compact radio sources at $22 \; \mathrm{GHz}$ and $43 \; \mathrm{GHz}$ \citep{vanderTak2005c}.  
The blue-filled star marks the position of the high-mass star, IRS7, powering the hypercompact \HII region, W3~F, while another infrared source, IRS~N7 (blue-open star), without any radio counterpart, is also shown \citep{Bik2012}.  
Several outflow axes (and a bipolar outflow along line of sight) reported in the literature are presented with labels of their associated tracers, including molecular outflows \citep[e.g.][]{Mitchell1991,Rodon2008,Wang2012}, water masers \citep{Imai2000}, and radio compact sources \citep{Wilson2003}.    
The gray dashed contours outline the more extended $6 \; \mathrm{cm}$ radiation at $4 \; \mathrm{mJy \, beam^{-1}}$ with $3^{\prime\prime}$ resolution. 
(b) The SMA polarized intensity map (color scale) overlaid with the orientation of the projected magnetic field, $B_\mathrm{pos}$ (line segment), which corresponds to dust polarization orientation rotated by $90^\circ$. 
Line segments are sampled twice per synthesized beam in locations where $I > 3\sigma_I$ and $P > 3\sigma_P$ (in black), and also $2\sigma_P < P < 3\sigma_P$ (in purple). 
The length of line segments is proportional to the fractional linear polarization in percentage, $P/I$, and the scalebar is shown in the upper right.    
Three thick line segments in gray show the large-scale magnetic field orientation measured from single-dish polarimeters. 
Red and blue arrows indicate bipolar outflows along with the outflow label. 
\label{fig:bvec}} %  
\end{figure}

Fig.~\ref{fig:bvec}b shows the polarized intensity map (color scale) overlaid with orientations of the projected magnetic fields, $B_\mathrm{pos}$ (line segments), which correspond to dust polarization orientations rotated by $90^\circ$.  
The overall geometry of the magnetic fields in our map is remarkably similar to the field pattern in the TADPOL map with a slightly larger beam size ($2.9^{\prime\prime}$) at a lower frequency \citep{Hull2014}.   
The orientations of $B_\mathrm{pos}$ appear organized in two parts: an hour-glass morphology centered at SMM2 (blue cross at the continuum peak) and a concave shape centered at IRS7 (blue star).
The symmetry axis of the hour-glass geometry is roughly aligned with the large-scale magnetic field orientation (thick gray segments) observed with single-dish polarimeters, including Hertz at $350 \;\mu\mathrm{m}$ \citep{Dotson2010} and SCUBA at $850 \; \mu\mathrm{m}$ \citep{Matthews2009}.  
The southern part shows a fairly organized concave morphology, possibly resulting from the expansion of the W3~F UC\HII region, ionized by IRS7, a spectral type B0--O9.5 ZAMS star \citep{Wynn-Williams1972,Tieftrunk1997,Bik2012}.
From the H$66\alpha$ recombination line data, \citet{Tieftrunk1997} found a velocity difference of roughly $5 \; \mathrm{km \, s^{-1}}$ across W3~F and suggested a scenario of a bow shock created by IRS7 moving toward the dense molecular core surrounding W3~IRS5.  
The bright limb on the northern edge of W3~F indicates an ionization front caused by the ram pressure of IRS~7. Meanwhile, the thermal pressure from the high electron gas temperature, around $10^4 \; \mathrm{K}$, generates a champagne flow toward the lower-density gas in the south and exhibits a cometary shape. 
%% The "ht!" tells LaTeX to put the figure "here" first, at the "top" next
%% and to override the normal way of calculating a float position
%% Fig. 1

%%%%%%%%%%%%%
\subsection{Molecular Outflows \label{sec:cosio}} %
W3~IRS5 has been known to drive multiple outflows observed with various tracers (Fig.~\ref{fig:bvec}a), including molecular emissions \citep[e.g.][]{Mitchell1991,Rodon2008,Wang2012}, proper motions of water masers \citep{Imai2000} and compact radio knots \citep{Wilson2003}.  
Figure~2 presents integrated intensity maps of the $\mathrm{CO \; (3-2)}$ and $\mathrm{SiO \; (8-7)}$ emissions and the corresponding position-velocity (PV) diagram cut along the outflow axis labeled 'b' and 'd.' 
%Meanwhile, a strong bipolar outflow, referred to as outflow-c, lying close to the line of sight, has also been detected in the center.  
The outflow labels follow those used in \citet{Wang2012}. 
The outflows are more discernible in the $\mathrm{CO \; (3-2)}$ data, possibly due to the higher excitation energies.
Overall, the CO emission exhibits a significantly broader range of velocities, $\Delta v =  -50$ to $100 \; \mathrm{km \, s^{-1}}$ (relative to the systemic velocity of $v_\mathrm{LSR} = -39.0 \; \mathrm{km \, s^{-1}}$), compared to the SiO emission. 
The red-shifted flow of $\mathrm{CO \; (3-2)}$ also has a broader range of velocities compared to that of the $\mathrm{CO \; (2-1)}$ in previous studies \citep{Wang2012}. 
The most prominent outflow in our data is outflow-b, which displays a well-collimated bipolar morphology along the east-west direction in both the CO and SiO emissions.
Yet outflow-b has only been detected in molecular gas and not associated with other outflow tracers, such as water masers or radio continuum knots.  
The PV diagrams of outflow-b show typical jet-driven structures where the maximum velocity increases with distance from the central source \citep[i.e. the ``Hubble wedges,'';][]{Arce2002} and a convex spur-like feature with the largest velocity spread at the farthest distance \citep{Masson1993,Lee2001}.
Notably, a compact CO knot is identified at the farthest position $(-9\farcs9,1\farcs0)$ of the red-shifted flow with a velocity offset reaching up to $\Delta v = 100 \; \mathrm{km \, s^{-1}}$.
The dynamical timescale for this redshifted CO knot is roughly 820~yr.     
The SiO emission also shows a compact knot closer to the central source with a velocity offset $\Delta v = 16 \; \mathrm{km \, s^{-1}}$.   
For outflow-d, the red-shifted flow toward the south dominates the emissions, while the blue-shifted flow toward the north is barely detected in the SiO emission.   
The $\mathrm{SiO \; (2-1)}$ emission also reports such asymmetric structure, possibly attributed to asymmetric structures in the ambient clouds.    
In all the PV diagrams, a huge velocity spread is observed at the center, suggesting a bipolar outflow with its axis lying close to the line of sight, referred to as outflow-c.  
This bipolar outflow may have an extensive range of velocities from $\Delta v = -63$ to $80 \; \mathrm{km \, s^{-1}}$.  
However, without spatial shifts, the confusion with numerous $\mathrm{^{34}SO_2}$ lines makes it challenging to identify plausible knotty structures for outflow-c.  
For example, the emission feature at velocity $62 \; \mathrm{km \, s^{-1}}$ should be the $\mathrm{^{34}SO_2 \; (4_{4,0}-4_{3,1})} \; v=0$ line emission.    
We did not detect outflow-a along NW-SE, possibly due to its extended spatial distribution \citep{Mitchell1991}.  

%% Fig. 2
\begin{figure}[h!] %
\epsscale{1.1} %
\plotone{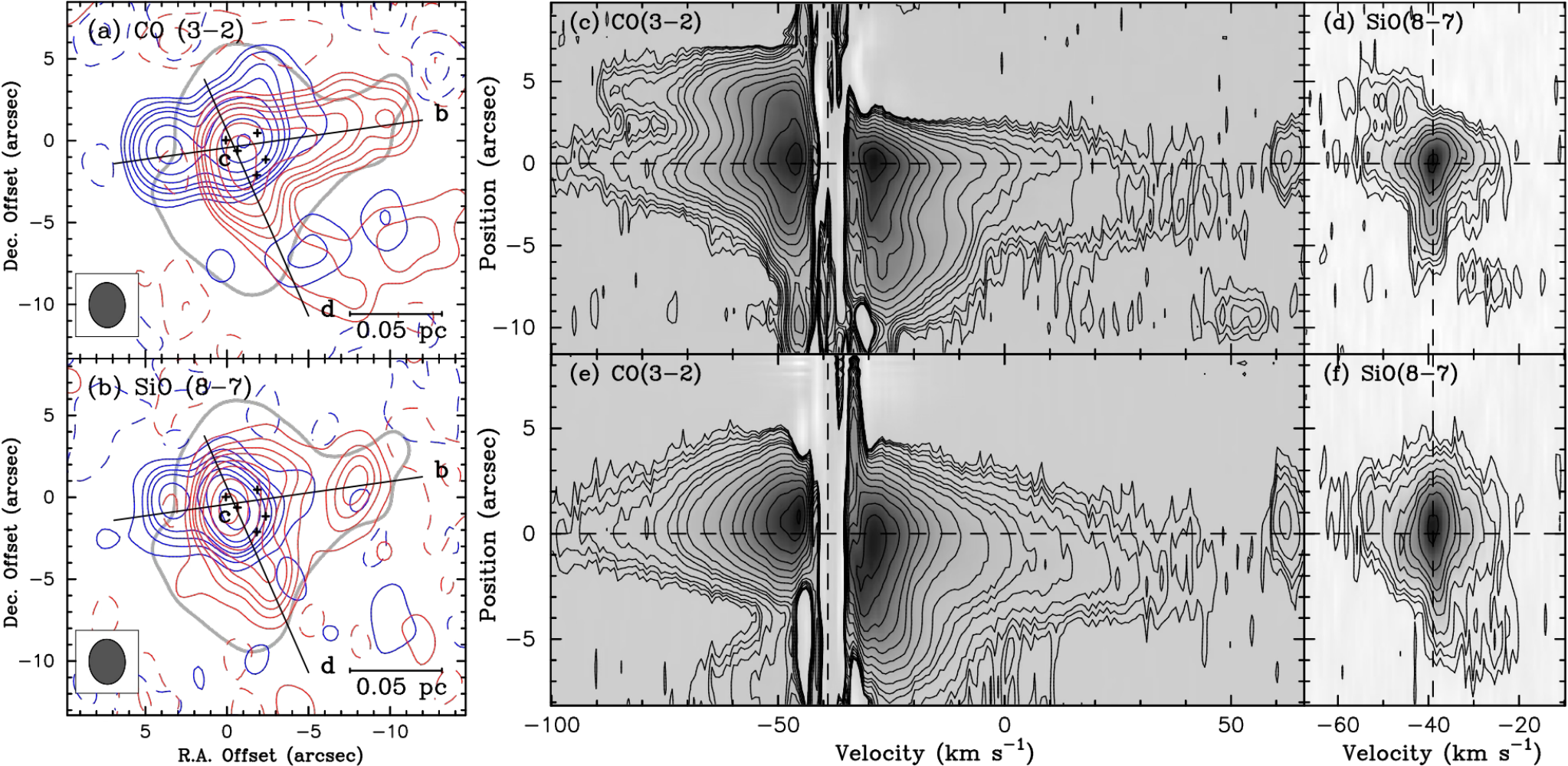} %
\caption{(a-b) Integrated intensity maps of the $\mathrm{CO \; (3-2)}$ and $\mathrm{SiO \; (8-7)}$ emissions overlaid on the VLA $6 \; \mathrm{cm}$ continuum map \citep[thin gray contours; ][]{Tieftrunk1997}.
The blue and red contours represent the blue-shifted and red-shifted components, respectively.  
Black crosses mark the peak positions of the five continuum sources \citep{Wang2013}. 
The thick gray contour outlines the $340 \; \mathrm{GHz}$ continuum emission at the $3\sigma$ level. 
Solid lines with labels show the PV cuts present in panel (c-f).  
Label c indicates the bipolar outflow lay close to the line of sight in the center.  
For the CO emission in panel (a), the blue-shifted component is integrated over a velocity range of $-90$ to $-62 \; \mathrm{km \, s^{-1}}$ with contours plotted at $(-12,-6,6,12,24,36,60,100,160,240) \times \sigma$, where $\sigma = 0.91 \; \mathrm{K \, km \,s^{-1}}$. 
The red-shifted component is integrated from $-15$ to $61 \, \mathrm{km \, s^{-1}}$ with contours at $(-12,-6,6,12,24,36,60,100,160) \times \sigma$, where $\sigma = 1.51 \; \mathrm{K \, km \, s^{-1}}$. 
For the SiO emission in panel (b), the blue-shifted component is integrated from $-55$ to $-43 \; \mathrm{km \, s^{-1}}$ with contours at $(-6,-2,2,6,10,14,22,30,50) \times \sigma$, where $\sigma = 0.60 \; \mathrm{K \, km \,s^{-1}}$. 
The red-shifted component is integrated from $-36$ to $-22 \, \mathrm{km \, s^{-1}}$ with contours plotted at $(-6,-2,2,6,10,14,22,30,50,70) \times \sigma$, where $\sigma = 0.65 \; \mathrm{K \, km \, s^{-1}}$. 
The beam sizes are $2\farcs7 \times 2\farcs2$  ($\mathrm{P. A.} = 2^\circ$). 
(c-f) Position-velocity diagram sliced along cut b (c--d) and along cut d (e--f) for the $\mathrm{CO \; (3-2)}$ and $\mathrm{SiO \; (8-7)}$ cubes.  
The systemic velocity of $v_\mathrm{LSR} = -39.0 \; \mathrm{km \, s^{-1}}$ and the center position located at SMM2 are marked by vertical and horizontal dashed lines.
Contours are at $(2,4,6,10,18,26,34,50,80,120,170,230,300,400,500,600) \times \sigma$, where $\sigma = 0.19 \; \mathrm{K}$. 
\label{fig:cosio}} %
\end{figure} %

\subsection{Tentative Velocity Gradients in the Blended $\mathrm{H^{13}CN} \; (4-3)$ and $\mathrm{SO_2 \; (13_{2,12}-12_{1,11})}$ Emissions \label{sec:vgrad}} %
\begin{figure}[h!] %
\epsscale{0.9} %
\plotone{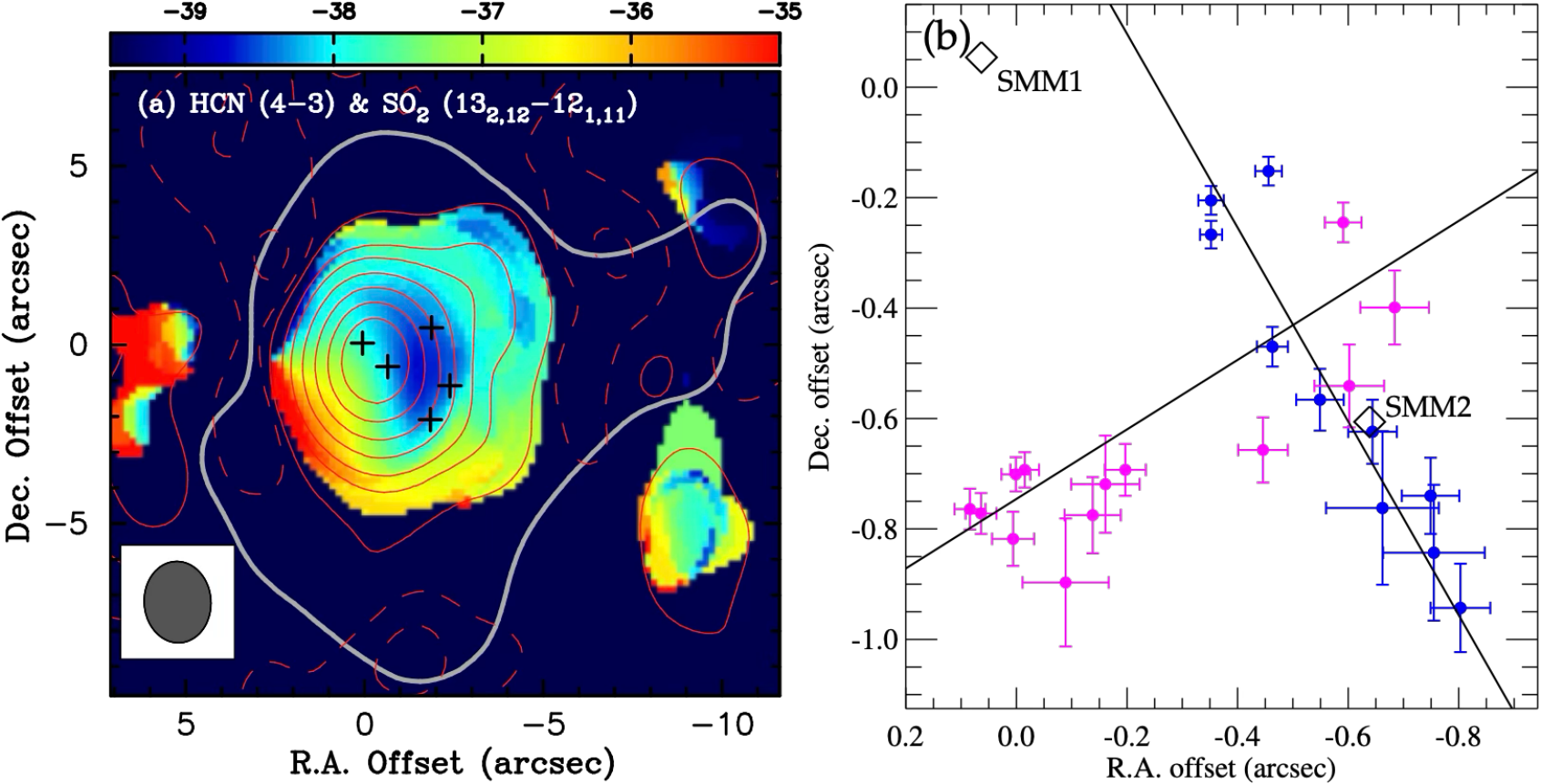} %
\caption{
(a) Integrated intensity map of the $\mathrm{H^{13}CN \; (4-3)}$ and $\mathrm{SO_2 \; (13_{2,12}-12_{1,11})}$ emissions (red contours) overlaid on intensity-weighted velocity map (color scale).   
The $\mathrm{H^{13}CN}$ line is blended with the $\mathrm{SO_2}$ line, which is redshifted by $1.06 \; \mathrm{km \, s^{-1}}$ relative to the $\mathrm{H^{13}CN}$ line.  
Red contours are plotted at $(-12,-3,3,20,40,80,140,220,320) \times \sigma$, where $\sigma = 1.4 \; \mathrm{K \, km \, s^{-1}}$.
Crosses mark the peak positions of the five continuum sources \citep{Wang2013}.  
The thick gray contour outlines the $340 \; \mathrm{GHz}$ continuum emission at the $3\sigma$ level.  
The beam is shown in the lower-left corner with a size of $2\farcs3 \times 1\farcs9$ (P.A. $=3^\circ$). 
(b) Peak positions of the 2D Gaussian fits for channels with emission above $6\sigma$, where $\sigma = 0.13 \; \mathrm{Jy \, beam^{-1}}$.  
Black lines represent the linear regressions to peak positions with uncertainties.  
The derived P.A. $= 30^\circ$ and $122^\circ$ for data in the velocity range of $-47.0$ to $-39.8 \; \mathrm{km \, s^{-1}}$ (blue) and $-39.0$ to $-29.4 \; \mathrm{km \, s^{-1}}$ (red), respectively.  
The velocity increases from south to north (blue) and then from west to east (red).  
\label{fig:h13cn}} %
\end{figure}

We investigate all the emission lines available in the passbands to probe rotation motions.  
Although W3~IRS5 are conspicuous in sulfur-bearing molecular lines \citep[e.g. ][]{Helmich1997}, many bright lines show extended structures, which unavoidably suffer from the lack of short spacing data and are excluded from the kinematics studies.    
For lines without severe missing flux issues, the intensity-weighted velocity maps show no clear rotation motion over the entire core.  
We further select emission lines with singly peaked morphology and perform 2D Gaussian fits for channels with emission above $6\sigma$ to determine the peak positions channel by channel for identifying any ordered kinematics.   
Velocity gradients along two distinct directions are discovered in the $\mathrm{H^{13}CN \; (4-3)}$ line (Fig.~\ref{fig:h13cn}).
The peak position moves from south to north (blue) along a line at $\mathrm{P.A.} = 30^\circ$ with velocity from $-47.0$ to $-39.8 \; \mathrm{km \, s^{-1}}$, and then moves from west to east (red) along a line at $\mathrm{P.A.} = 122^\circ$ with velocity from $-39.0$ to $-29.4 \; \mathrm{km \, s^{-1}}$.  
The direction of the aligned peak positions is determined by a linear fit to peak positions with uncertainties in both the right ascension and declination.  
However, the velocity gradients cannot be determined quantitatively due to the confusion caused by line blending.  
The $\mathrm{H^{13}CN \; (4-3)}$ line at $\nu_0 = 345.33976 \; \mathrm{GHz}$ is blended with the $\mathrm{SO_2 \; (13_{2,12}-12_{1,11})}$ line at $\nu_0 = 345.33854 \; \mathrm{GHz}$, which is roughly redshifted by $1.06 \; \mathrm{km \, s^{-1}}$ relative to the $\mathrm{H^{13}CN}$ line.  
The upper-level energies of the two blended lines are quite similar, $E_\mathrm{up} = 41 \; \mathrm{K}$ and $93 \; \mathrm{K}$ for $\mathrm{H^{13}CN \; (4-3)}$ and $\mathrm{SO_2} \; (13_{2,12}-12_{1,11})$, respectively.  

Velocity gradients in two distinct directions nearly perpendicular to each other have been reported by \citet{Wang2013} in the $\mathrm{SO_2 \; (34_{3,31}-34_{2,32})}$ and $\mathrm{HCN \; (4-3)} \, v_2 = 1$ emissions, where $E_\mathrm{up} = 587$ and $1067 \; \mathrm{K}$, respectively.   
We reanalyze their data and improve the image quality of the $\mathrm{SO_2}$ cube (see Appendix~\ref{sec:so2}).  
The peak position of the $\mathrm{SO_2 \; (34_{3,31}-34_{2,32})}$ emission moves along two directions at $\mathrm{P.A.} = 31^\circ$ and $132^\circ$, similar to the motions found in our data.  
Despite different excitation conditions, these spectral lines reveal velocity gradients along two distinct directions around SMM2. 
These velocity gradients are often related to unresolved kinematics of outflows or rotating disks.   
Since both HCN and $\mathrm{SO_2}$ can be present in outflows and disks, observations with higher angular resolutions are required to reveal the nature of the two velocity gradients. 

\iffalse %
The N-S has $\mathrm{P.A.} = 30^\circ$ and E-W has $\mathrm{P.A.} = 122^\circ$. 
The $\mathrm{H^{13}CN \; (4-3)}$ ($345.33976 \; \mathrm{GHz}$) emission is blended with the $\mathrm{SO_2 \; (13_{2,12}-12_{1,11})}$($345.33854 \; \mathrm{GHz}$), which is redshifted by $1.06 \; \mathrm{km \, s^{-1}}$ relative to the $\mathrm{H^{13}CN}$ line.  
The $\mathrm{H^{13}CN} \; (4-3)$ line ($\nu_0 = 345.3397599 \; \mathrm{GHz}$, $E_\mathrm{up} = 41.43495 \; \mathrm{K}$ and $\log A_{ij} = -2.69341$, $A_{ij} = 2.02577\times10^{-3}$).  
The $\mathrm{H^{13}CN}$ line is blended with $\mathrm{SO_2} \; (13_{2,12}-12_{1,11})$, approximately red-shifted by $\sim 1.06 \; \mathrm{km \, s^{-1}}$ ($\nu_0 =  345.3385391 \; \mathrm{GHz}$, $E_\mathrm{up} = 92.98367 \; \mathrm{K}$ and $\log A_{ij} = -3.62328$, $A_{ij} = 2.380784\times10^{-4} = 0.117525 \times A_{ij}^\mathrm{H^{13}CN}$) 
\fi %

%% Fig. 3

\section{Analysis \label{sec:analysis}} % 
\subsection{Core Mass Estimate \label{sec:mass}} %
Our continuum map has a similar spatial coverage and comparable shortest baselines to the 223~GHz continuum map from the TADPOL/CARMA survey \citep{Hull2014}.  
This allows us to derive the spectral index by comparing the flux density between 223~GHz and 340~GHz.    
Excluding the western bridge possibly affected by the free-free emission of W3~B, we measure the flux density of the IRS~5 core to be 1.2~Jy and 6.4~Jy at 223.4~GHz and 340.0~GHz, respectively, and derive a spectral index of $4.0$, rendering a dust opacity spectral index of $\beta = 2.0$. 
%The 25\% uncertainty is caused by the 15\% flux measurement uncertainty at the two observed frequencies. }
%Steep spectral indices, $\beta \gtrsim 2$, have also been reported by \citet{Shirley2011} with well-constraint measurements at $850$ and $450 \; \mu\mathrm{m}$.   
The core mass, $M_g$, can be estimated from the measured flux density, $F_\nu$, at the observing frequency, $\nu$, with 
\begin{equation} %
M_g = {\cal{R}_\mathrm{gd}} \frac{ F_\nu d^2 \,  }{\kappa_\nu B_\nu(T_d)},
\end{equation} %
where $d$ is the distance to the source, $\kappa_\nu$ the dust opacity, and $B_\nu(T_d)$ the Planck function at a dust temperature, $T_d$.  
%For temperature measurements, fitting the spectral energy distribution (SED) obtained a dust temperature of $118 \; \mathrm{K}$ \citep{Palau2021} while analyzing excitation conditions of molecular lines obtains gas temperatures ranging from $90 - 160 \; \mathrm{K}$ \citep{Wang2013}.  
%We adopt a moderate value of $T = 120 \; \mathrm{K}$ for temperature in all our analyses.    
For the dust temperature, we adopt a value of $T = 120 \pm 14 \; \mathrm{K}$, which approximates the average dust temperature of $118 \; \mathrm{K}$ within the central spherical volume of $0.15 \; \mathrm{pc}$ derived from the spectral energy distribution with an uncertainty of 12\% \citep{Palau2021}. 
This temperature value also falls within the gas temperature range of $90 - 160 \; \mathrm{K}$, as determined by analyzing the excitation conditions of molecular lines \citep{Wang2013}.
Applying the widely used value of the canonical gas-to-dust mass ratio, ${\cal{R}_\mathrm{gd}} = 100$, and a dust opacity of $\kappa_\nu = 10 (\lambda/250 \; \mu\mathrm{m})^{-\beta} = 0.8 \; \mathrm{cm^2 \, g^{-1}}$ \citep{Hildebrand1983}, we estimate the gas mass of the IRS~5 core to be $M_g = 32 \, M_\odot$.   
The main uncertainty in mass determination arises from the poorly characterized ${\cal{R}_\mathrm{gd}}$ and $\kappa_\nu$.  
Following the discussion by \citet{Sanhueza2017}, we adopt the uncertainty of 23\% for ${\cal{R}_\mathrm{gd}}$ and 19\% for $\kappa_\nu$ derived from the measurements at $850 \mu\mathrm{m}$ by \citet{Shirley2011}.
Including all the error estimates, the gas mass uncertainty is 39\%.     
The deconvolved size of the continuum emission yields a core radius of $R = 0.051 \pm 0.004 \; \mathrm{pc}$. 
Assuming a spherical structure, we obtain the gas mean density of $n_\mathrm{H_2} = M_g/(4\pi R^3/3) =  10^6 \; \mathrm{cm^{-3}}$ and the mean column density of $N_\mathrm{H_2} = M_g/\pi R^2 = 2.1 \times 10^{23} \; \mathrm{cm^{-2}}$.  
The uncertainty of the gas mean density is 45\%, and that of the mean column density is 42\%.
In addition, the ionized flux densities produced by hypercompact \HII regions, d1/d2 (SMM1) and b (SMM2), suggest these protostars to be spectral type B0.5-B1 \citep{Wilson2003}, equivalent to a stellar mass of $11 \; M_\sun$ for each source \citep{Hanson1997}. 
Including both the gas mass, $M_g = 32 \, M_\odot$, and the sum of stellar mass, $M_* = 22 \, M_\odot$, the total mass related to gravitational energy is $M = M_g + M_* = 54 \; M_\odot$, of which 41\% is in protostars.

\subsection{Modeling the Magnetic Field} % 
The magnetic field model consists of two parts: an hourglass model centered at SMM2 and a uniform spherical model centered at IRS7. 
We use the Spheroid Flux Freezing (SFF) model developed by \citet{Myers2018} to describe the hourglass geometry.    
The SFF model is based on a Plummer spheroid that has contracted under self-gravity from an initially uniform magnetized medium while conserving flux, mass, and shape (axial ratios).  
%Based on the observed polarization pattern in W3~IRS5,
Because W3~IRS5 does not show linearly aligned multiple fragments, we choose a parallel oblate spheroid with axis ratios of $A=B > C=1$, a Plummer power-law index of $p=2$, and its symmetry axis aligned with the initial magnetic field.  
We use 4 parameters to specify the hourglass model: the position angle, $\phi_\mathrm{sff}$, and inclination angle (relative to the line of sight), $i_\mathrm{sff}$, of the symmetry axis, the axis ratio of an oblate spheroid, $A_\mathrm{sff}$, and the normalized peak spheroid density,  $\rho_p \equiv n_p/n_\mathrm{bg}$, which is the peak spheroid density, $n_p$, normalized to the background value, $n_\mathrm{bg}$.    
The spherical model for the perturbation in the southern region is empirical, featuring a uniform magnetic field strength, $b_\mathrm{sph}$, that consists only of an azimuthal component within an effective radius, $r_\mathrm{sph}$.
%The parameter $b_\mathrm{sph}$ is a scaling factor adjusting the relative importance between the two model components.  
The edge of the spherical model is tapered by the complementary error function with a width of $0.2 \, r_\mathrm{sph}$.  
The spherical model is placed at the same distance as the hourglass model with its polar axis aligned with the line of sight.  
The two magnetic field models are combined using vector addition to determine the direction of the local magnetic field. 
There are 6 parameters to describe the final magnetic field geometry in our models: $\phi_\mathrm{sff}$, $i_\mathrm{sff}$, $A_\mathrm{sff}$, $\rho_p$, $r_\mathrm{sph}$, and $b_\mathrm{sph}$.  

\begin{figure}[h!] %
\epsscale{1.15} %
\plotone{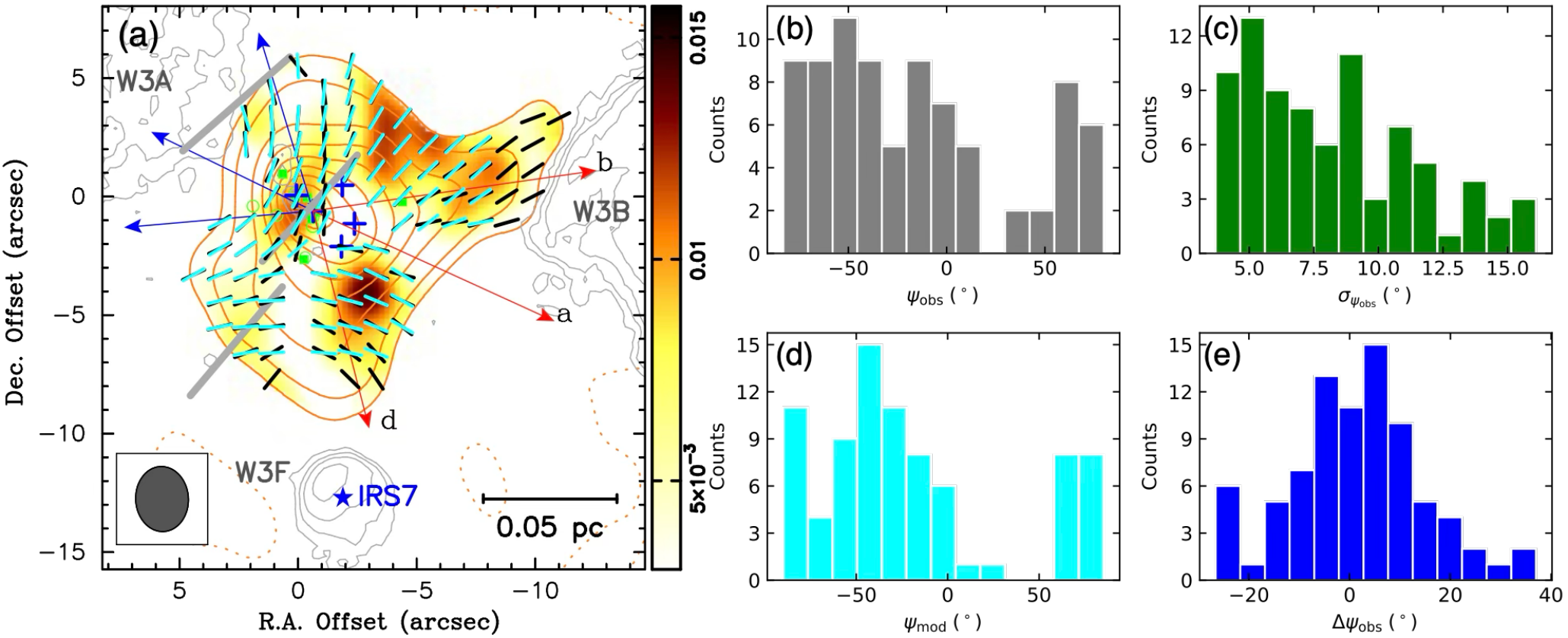} %
\caption{(a) Best-fit magnetic field model (cyan) overlaid on the observed magnetic field orientation (black) for models composed of an hourglass component centered at SMM2 combined with a spherical component centered at IRS~7.  
Data on the outskirts are excluded for optimization.  
The length of line segments is fixed and arbitrary.  
Red and blue arrows indicate bipolar outflows, same as the outflow axes shown in Fig.~\ref{fig:bvec}(a).
Symbols are the same as those in Fig.~\ref{fig:bvec}(b).
(b)--(e) Histogram of anglar distribution for (b) observed position angle, $\psi_\mathrm{obs}$, (c) measurement uncertainty, $\sigma_{\psi_\mathrm{obs}}$, (d) position angle of the best-fit model, $\psi_\mathrm{mod}$, and (e) field oritation residual, $\Delta \psi_\mathrm{obs} = \psi_\mathrm{obs} - \psi_\mathrm{mod}$.  
\label{fig:bmod}} %
\end{figure}

To interpret the observed polarization pattern in W3 IRS~5, we calculate the Stokes~$I$, $Q$, and $U$ model images for a given magnetic field geometry by solving for 3D radiative transfer of the polarized dust emission arising in the envelope component as described in \citet{Chen2016b}.  
Consider a local magnetic field direction, $\vec{\hat{b}}$, at an inclination angle, $i_b$, relative to the line of sight, $\vec{\hat{s}}$, with the projected component oriented at a position angle, $\psi_b$, on the plane of the sky.  
Following \citet{Draine2022}, the radiative transfer equations for the Stokes parameters are given by (see Appendix~\ref{sec:iqueqn}) 
\begin{eqnarray} %
\frac{\dif I}{\dif s} &=& \rho \kappa_\nu \left( -\left[1+\alpha \left( \cos^2 i_b -\frac{1}{3} \right) \right] [I-B_\nu(T_d)] + \alpha \sin^2 i_b \left[ \cos 2\psi_b Q + \sin 2\psi_b U \right] \right) \label{eq:rti} \\
\frac{\dif Q}{\dif s} &=& \rho \kappa_\nu \left( \alpha \cos 2\psi_b \sin^2 i_b [I - B_\nu(T_d)] - \left[1+\alpha \left( \cos^2 i_b -\frac{1}{3} \right) \right] Q \right) \label{eq:rtq} \\
\frac{\dif U}{\dif s} &=& \rho \kappa_\nu \left( \alpha \sin 2\psi_b \sin^2 i_b [I - B_\nu(T_d)] - \left[1+\alpha \left( \cos^2 i_b -\frac{1}{3} \right) \right] U \right), \label{eq:rtu} %
\end{eqnarray} %
where $\alpha \equiv f_\mathrm{align} C_\mathrm{pol}/C_\mathrm{ran}$ is the ratio of the cross section for an aligned grain ensemble to that of a randomly oriented grain ensemble.  
We assume a constant value for $\alpha$ and estimate $\alpha = 0.17$ using the maximum degree of polarization $p_\mathrm{max} = 18\%$ in our data.  
After calculating the Stokes~$I$, $Q$, and $U$ model images, we perform synthetic imaging using MIRIAD for the SMA observations of these model Stokes images.
We then compare the resulting position angles of the magnetic field orientations with those observed in the data.   
To produce more extended continuum emission, we include six compact sources with their associated envelopes:  the five compact submillimeter sources \citep[blue crosses;][]{Wang2013} and the radio source K1 \citep[green square at position offset $(0^{\prime\prime},-4^{\prime\prime})$;][]{vanderTak2005c}.  
Based on previous continuum observations, the stellar masses are assumed to be $12, 10, 6, 6, 4, 5 \, M_\odot$.     
The Levenberg-Marquardt method is employed to optimize magnetic field models by minimizing the $\chi^2$ value, which is calculated by comparing the model P.A. with the observed values in 82 positions.
The optimized model (Fig.~\ref{fig:bmod}a) gives a minimum reduced $\chi^2$ value of $\overline{\chi^2}_\mathrm{min}= 3.55$.  
%For comparison, we also show the magnetic field geometry of the optimized model without the spherical component in the south (Fig.~\ref{fig:bmod}b).      
The parameters of the best-fit model are listed in Table~\ref{tab:bestfit} with uncertainties estimated from the confidence region within $\Delta \chi^2 = 7.04$, corresponding to the 68.3\% confidence level with 6 parameters \citep[see Chapter 15,][]{Press1992}. 
The Stokes-$I$, $Q$, and $U$ images for the optimized model are shown in Fig.~\ref{fig:modiqu}.

We consider the observed pinched morphology of magnetic fields to have resulted from a gravitational contraction of a clump threaded by an initially uniform background magnetic field seen at large scales.   
The contraction increases the peak density from the background value by a factor of $\rho_p = 9.7 \times 10^3$.  
Applying the same background density of $n_\mathrm{bg} = 300 \; \mathrm{cm^{-3}}$ in the SFF model \citep{Myers2018}, the best-fit model implies a peak density of $n_p = 2.9 \times 10^6 \; \mathrm{cm^{-3}}$, similar to what we have derived from the continuum emission.  

%The projected magnetic fields ($B_\mathrm{pos}$) show complicated patterns at clump scales and do not concur with one single large-scale ($\sim 1^\prime$) hourglass model at $\mathrm{P.A. \sim 50^\circ}$ based on early observational results, such as the large-scale CO outflow \citep{Mitchell1991}, a gradient of the magnetic field along the line of sight \citep[$B_\mathrm{los}$;][]{Roberts1993}, and scarcely sampled single-dish measurements \citep{Greaves1994}.  

The axis of the SSF component lies very close to the clump-scale (background) magnetic field at $\mathrm{P.A.} \sim 140^\circ$ measured at clump scales \citep[$\sim 20^{\prime\prime}$ or $0.18 \; \mathrm{pc}$;][]{Matthews2009,Dotson2010}.  
This orientation, however, is different from the orientation of a single large-scale ($\sim 1^\prime$ or $0.5 \; \mathrm{pc}$) hourglass model at $\mathrm{P.A. \sim 50^\circ}$ based on early observational results, such as a gradient of the magnetic field along the line of sight, $B_\mathrm{los}$, derived from \HI Zeeman effects \citep{Roberts1993} and scarcely sampled single-dish dust polarization measurements for $B_\mathrm{pos}$  \citep{Greaves1994}.  
Those early polarimetric studies were conducted over the entire W3~Main cloud, and most likely suffered from source confusion with all the nearby \HII regions.  
However, the axis of the hourglass model that we proposed here at core scales ($\sim 0.05 \; \mathrm{pc}$) does not align with the large-scale CO outflow \citep[{\bf $88^\circ$ apart;}][]{Mitchell1991} but appears closer to the axis of outflow-b ($57^\circ$ apart), which drives the compact CO knot with the largest velocity offset of $\Delta v = 100 \; \mathrm{km \, s^{-1}}$ (Sect.~\ref{sec:cosio}).  
 
Once the optimized model is found, we compare the observed field orientation, $\psi_\mathrm{obs}$, with the best-fit model value, $\psi_\mathrm{mod}$, to calculate the field orientation residuals, $\Delta \psi_\mathrm{obs} = \psi_\mathrm{obs} - \psi_\mathrm{mod}$.  
These residuals, $\Delta \psi_\mathrm{obs}$, are measurement uncertainties, $\sigma_{\psi_\mathrm{obs}}$, convolved with the intrinsic angular dispersion, $\delta \psi$, which is caused by Alfv\'{e}n waves propagating along magnetic field lines and can be used to probe the magnetic field strength (see Sect.~\ref{sec:Bfield}). 
The distributions of $\psi_\mathrm{obs}$, $\psi_\mathrm{mod}$, $\sigma_{\psi_\mathrm{obs}}$, and $\Delta \psi_\mathrm{obs}$ are shown in Fig.~\ref{fig:bmod}(b)--\ref{fig:bmod}(e).  
We first compute the standard deviation of $\Delta \psi_\mathrm{obs}$ to obtain the observed dispersion, $\delta \psi_\mathrm{obs} = 13.7^\circ$.      
This observed dispersion, $\delta \psi_\mathrm{obs}$, includes the measurement uncertainty dispersion, $\delta \sigma_{\psi_\mathrm{obs}} = 8.9^\circ$, which is the root mean square (rms) of the measurement uncertainties, $\sigma_{\psi_\mathrm{obs}}$.  
The intrinsic angular dispersion can be calculated from $\delta \psi = \sqrt{\delta \psi_\mathrm{obs}^2 - \delta \sigma_{\psi_\mathrm{obs}}^2} = 10.4^\circ$.  
To estimate the uncertainty in $\delta \psi$, we generate a bootstrapping statistical sample of 500,000 synthetic data sets, each of which assumes a Gaussian distribution centered at the orientation residuals, $\Delta \psi_\mathrm{obs}$, with the corresponding measurement uncertainty, $\sigma_{\psi_\mathrm{obs}}$.  
Care has been taken to ensure a sufficient sample size for convergence.  
We then calculate the standard deviation of the intrinsic angular dispersion derived from the bootstrapping samples and obtain the uncertainty in $\delta \psi$ to be $1.2^\circ$, roughly $12\%$.  
The resulting intrinsic polarization angle dispersion is $\delta \psi = 10.4^\circ \pm 1.2^\circ$.  

\begin{deluxetable}{lc} %
%\tablenum{1}
\tablecaption{Parameters of the Optimized Model\label{tab:bestfit}}
\tablewidth{0pt}
\tablehead{
\colhead{Parameter} & \colhead{Value} } %
%\decimalcolnumbers
\startdata
\multicolumn{2}{l}{Hourglass component: SFF model} \\
\cline{1-1} %
Position angle, $\phi_\mathrm{sff}$ & $152^\circ \pm 5^\circ$ \\
Inclination angle, $i_\mathrm{sff}$ & $38^\circ \pm 11^\circ$ \\
Axis ratio, $A_\mathrm{sff}$ & $4.1 \pm 0.6$ \\
Normalized peak spheroid density, $\rho_p$ & $(9.7 \pm 0.9) \times 10^3$ \\
\hline %
\multicolumn{2}{l}{Spherical component} \\
\cline{1-1} %
Effective radius, $r_\mathrm{sph}$ & $10\farcs2 \pm 0\farcs5$ \\
Scaling factor, $b_\mathrm{sph}$ & $7 \pm 1$ \\
\enddata
%\tablecomments{} %
\end{deluxetable}

%% Fig. 4

\subsection{Magnetic Field Strength \label{sec:Bfield}} %
With the intrinsic dispersion of the field orientations, $\delta \psi$, we apply the Davis-Chandrasekhar-Fermi \citep[DCF;][]{Davis1951,Chandrasekhar1953} method to estimate the field strength of the projected magnetic field component, i.e. the plane-of-sky component, $B_\mathrm{pos}$.  
The DCF method considers the dispersion in field orientations, $\delta \psi$, from the local projected fields as observable for the transverse motion of Alfv\'{e}n waves.   
However, the transverse velocity dispersion cannot be measured observationally.   
If assuming an isotropic velocity dispersion, one may approximate the transverse velocity dispersion with the observed Gaussian line width, $\delta v_\mathrm{los}$, which is the velocity dispersion along the line of sight. 
Together with the mass density, $\rho$, one can estimate the projected magnetic field strength with
 approximate the transverse velocity dispersion equal to \begin{eqnarray} %
B_\mathrm{pos} &=& Q \sqrt{4\pi \rho} \, \frac{\delta v_\mathrm{los}}{\delta \psi} = Q \sqrt{4\pi \mu_\mathrm{H_2} m_\mathrm{H} n_\mathrm{H_2}} \,  \frac{\delta v_\mathrm{los}}{\delta \psi}, \\
        &=& 6.342 \; \mathrm{mG} \left( \frac{Q}{0.5} \right) \left( \frac{\mu_\mathrm{H_2}}{2.33} \right)^{1/2} \left( \frac{n_\mathrm{H_2}}{10^5 \; \mathrm{cm^{-3}}} \right)^{1/2} \left( \frac{\delta v_\mathrm{los}}{1 \; \mathrm{km \, s^{-1}}} \right) \left( \frac{\delta \psi}{1^\circ} \right)^{-1},  
\end{eqnarray} %
where $Q$ is the correction factor, $\mu_\mathrm{H_2}$ the mean molecular weight per hydrogen molecule, $m_\mathrm{H}$ the mass of hydrogen atom, and $n_\mathrm{H_2}$ the gas number density.  
The correction factor, $Q$, accounts for the deviation between the actual and estimated magnetic field strength, and $Q=1$ corresponds to the original DCF formula. 
Using simulations for turbulent clouds, \citet{Ostriker2001} performed simulations with conditions in turbulent clouds and suggested $Q = 0.5$, applicable to $\delta \psi \lesssim 25^\circ$. 
In a later study, \citet{Liu2021} conducted a thorough analysis using the DCF method on numerical simulation data and suggested $Q = 0.33$ for uniform/ordered magnetic fields in spherical clumps with initial radii of $0.25$ and $0.5 \; \mathrm{pc}$.      
%The velocity dispersion measured with the $\mathrm{H^{13}CO^+ \; (4-3)}$ emission is $1.36 \; \mathrm{km \, s^{-1}}$ \citep{Palau2021}.  
To avoid overestimating the velocity dispersion, we carefully select line emission that does not trace outflow activities, especially the outflow-c lying close to the line of sight at the center \citep[Sect.~\ref{sec:cosio};][]{Wang2012}. 
The selected $\mathrm{CH_3OH \; (7_{1,7}-6_{1,6})}$ line emission shows a singly peaked morphology and a Gaussian line width of $\delta v_\mathrm{CH_3OH} = 0.90 \pm 0.03 \; \mathrm{km \, s^{-1}}$.  
The Gaussian line width consists of two components: thermal and nonthermal. 
The thermal component is related to the sound speed, $c_s = \sqrt{k T/m}$, where $m$ is the particle mass, while the nonthermal component corresponds to turbulent velocity dispersion,  $\delta v_t$.    
Assuming the gas and the dust have the same temperature of $T = 120 \; \mathrm{K}$, the sound speed is $c_s = 0.603 \; \mathrm{km \, s^{-1}}$ for gas of the average partcile mass, $\mu_\mathrm{H_2} m_\mathrm{H}$, and $c_{s,\mathrm{CH_3OH}} = 0.177 \; \mathrm{km \, s^{-1}}$ for $\mathrm{CH_3OH}$ molecules. 
The uncertainties in the sound speed estimates are 5.8\%.  
The turbulent velocity dispersion is computed with  
\begin{equation} %
\delta v_t = \sqrt{\delta v_\mathrm{CH_3OH}^2 - c_{s,\mathrm{CH_3OH}}^2}.  
\end{equation} %.   
Once the turbulent velocity dispersion is found, the gas velocity dispersion along the line of sight can be calculated with 
\begin{equation} %
\delta v_\mathrm{los} = \sqrt{c_s^2 + \delta v_t^2}.  
\end{equation} %
We calaculate a turbulent velocity dispersion of $\delta v_t = 0.88 \pm 0.03 \; \mathrm{km \, s^{-1}}$, which makes the turbulence transonic, $\delta v_t/c_s$ = 1.5, and leads to a gas velocity dispersion of $\delta v_\mathrm{los} = 1.07 \pm 0.03 \; \mathrm{km \, s^{-1}}$.  
Using $Q=0.33$, we estimate a projected magnetic field strength of $B_\mathrm{pos} = 1.36 \pm 0.35 \; \mathrm{mG}$.  

When Zeeman measurements are available to provide the line-of-sight component of the magnetic field strength, $B_\mathrm{los}$, it is possible to estimate the total field strength by combining the two components.     
The VLA observations for the Zeeman effect in the $\mathrm{H_2O}$ masers measured a field strength of $-18.6 \pm 1.3 \; \mathrm{mG}$ in the post-shocked region \citep{Sarma2002} .
Depending on the shock compression, the field can be amplified by a factor of roughly 20 \citep{Sarma2002}, rendering an estimate of the pre-shock field strength of $B_\mathrm{los} = -0.93 \pm 0.07 \; \mathrm{mG}$.
Hence, the total magnetic field strength can be calculated with $B_\mathrm{tot} = (B_\mathrm{pos}^2 + B_\mathrm{los}^2)^{1/2} = 1.6 \pm 0.4 \; \mathrm{mG}$.   
This value is roughly two times higher than the field strength of $B_0 \sim 0.7 \; \mathrm{mG}$ using the angular dispersion function (ADF) method on the TADPOL data, which cover a larger region of 0.4~pc (roughly two times larger than the SMA mapping area), including the \HII region, W3~B, in the west \citep{Houde2016}.  
The ADF method most likely extracts the ordered field strength on larger scales, corresponding to the magnetic fields shown by the single-dish observations, while our measurement is more related to the increased field strength affected by gravitational contraction.  
It is also useful to compare the maximum field strength predicted by the SFF model at the center of the pinched field morphology.  
Due to the flux freezing assumption in the SFF model, the field strength depends on the density contrast.   
Assming the same background field strength of $B_\mathrm{bg} = 10 \; \mu\mathrm{G}$ used in the SFF model \citep{Myers2018}, the maximum field strength will be $B_\mathrm{max,sff} = B_\mathrm{bg} (1 + \rho_p)^{2/3} = 4.5 \; \mathrm{mG}$, which is comparable but slightly higher than our estimate using the DCF method.  

\section{Discussion \label{sec:discussion}} % 
\subsection{The Normalized Mass-to-Magnetic-Flux Ratio} %
The magnetic field strength of $B_\mathrm{tot} = 1.6 \; \mathrm{mG}$ in W3~IRS5 is among typical values of a few mG reported in high-mass star-forming regions, e.g. G240.31+0.07 \citep{Qiu2014}, IRAS~18089$-$1732 \citep{Sanhueza2021}, G333.46$-$0.16 \citep{Saha2024}, IRAS~16547$-$4247 \citep{Zapata2024}.
The ratio of the mass in a magnetic flux tube to the magnitude of magnetic flux, $M/\Phi_B$, is an important parameter for ambipolar diffusion models \citep[e.g.][]{Nakano1978,Mouschovias1976,Li1996}.   
We calculate the ratio of the observed mass-to-magnetic-flux ratio, $M/\Phi_B$, normalized to its critical value, $(M/\Phi_B)_\mathrm{crit}$, to evaluate the stability of a clump 
\begin{eqnarray} %
\lambda &\equiv& \frac{(M/\Phi_B)}{(M/\Phi_B)_\mathrm{crit}} \\
&=& 2\pi \sqrt{G} \, \frac{M}{\pi R^2 B_\mathrm{tot}} = 1.08 \times 10^{-2} \left( \frac{M}{M_\odot} \right) \left( \frac{B_\mathrm{tot}}{1 \; \mathrm{mG}} \right)^{-1} \left( \frac{R}{0.1 \; \mathrm{pc}} \right)^{-2} \\
&=& 2\pi \sqrt{G} \, \frac{\mu_\mathrm{H_2} m_\mathrm{H} N_\mathrm{H_2}}{B_\mathrm{tot}} = 0.633 \left( \frac{\mu_\mathrm{H_2}}{2.33} \right) \left( \frac{N_\mathrm{H_2}}{10^{23} \; \mathrm{cm^{-2}}} \right) \left( \frac{B_\mathrm{tot}}{1 \; \mathrm{mG}}  \right)^{-1},
\end{eqnarray}
where $N_\mathrm{H_2}$ is the gas column density, and $(M/\Phi_B)_\mathrm{crit} = c_\Phi/\sqrt{G}$ is the critical value with $c_\Phi = 1/2\pi$ \citep{Nakano1978}.  
The criterion for gravitational contraction is $\lambda > 1$, that is, supercritical.
Yet the precise value of the numerical coefficient, $c_\Phi$, is slightly model dependent and may differ by a factor of $0.75 - 1.5$ \citep[e.g.][]{Strittmatter1966,Mouschovias1976,Tomisaka1988,Crutcher1999a}.   
For W3~IRS5, we find a normalized mass-to-magnetic-flux ratio of $\lambda = 0.9$ when only considering the gas mass $M_g = 32 M_\odot$.   
However, when we account for the stellar mass originating from the initial contracting clump, the total mass increases to $M = 54 M_\odot$ (see Sect.~\ref{sec:mass}), leading to $\lambda = 1.5$. 
The uncertainty in $\lambda$ is about 49\%. 
Considering the variation in $c_\Phi$ and the possibility of missing flux at scales beyond $12^{\prime\prime}$ ($\sim 0.1 \; \mathrm{pc}$), W3~IRS5 is moderately supercritical and is most likely collapsing.  

A mass-to-magnetic-flux ratio marginally supercritical ($\lambda \gtrsim 1$) is reported in several high-mass star-forming cores harboring a small OB group, e.g. G240.31+0.07 \citep{Qiu2014}, G31.41+0.31 \citep{Beltran2019,Beltran2021}, G333.46$-$0.16 \citep{Saha2024}.  
All these cores show a pinched magnetic field geometry with two or three main massive fragments, and occasionally accompanied by a few less massive ones.  
The presence of magnetic fields seems to effectively suppress fragmentation and lead to the formation of a small number of dense cores, consistent with magnetohydrodynamical (MHD) simulations for moderately supercritical cores \citep[e.g.][]{Commercon2011,Peters2011,Myers2013a}.   
Intriguingly, W3~IRS5 appears to be a rare case that does not show rotation at the core scale ($\sim 0.1~\mathrm{pc}$). 
It is not clear whether the magnetic field has been efficient in removing the angular momentum, or the collapse initially started with weak rotation.  
Alternatively, the strong multiple outflows and radiative feedback may perturb or disrupt the gas rotational motion.  

\subsection{Analysis of the Energy Balance and Virial Parameter} %
The relative importance of gravity, turbulence, and magnetic fields can be assessed by comparing their energy contributions.  
Since W3~IRS5 has a significant fraction of mass in the central protocluster (Sect.~\ref{sec:mass}), we consider the cluster as a point source located at the center of a clump with a power-law density distribution, $\rho \propto r^{-p}$,  and calculate the gravitational energy accounting for the stellar mass fraction of $s=0.41$ (see Appendix~\ref{sec:mvir}).  
The density structure of this core can be modeled with $p=1.46 \pm 0.04$ \citep{Palau2021}, which is obtained by simultaneously fitting the radial intensity profiles at $450$ and $850 \; \mu\mathrm{m}$ SCUBA images along with the spectral energy distribution \citep{Palau2014}.   
We compute the gravitational energy, $E_G = -4.6 \times 10^{45} \; \mathrm{erg}$, the kinetic energy, $E_K = 1.1 \times 10^{45} \; \mathrm{erg}$, and the magnetic energy, $E_B = 1.7 \times 10^{45} \; \mathrm{erg}$, using Eq.~(\ref{eq:EG}), (\ref{eq:EK}), and (\ref{eq:EB}), respectively.  
We explore the uncertainties associated with energy estimates by generating the corresponding bootstrapping statistics samples.  
Each of the parameters is sampled with a Gaussian distribution centered at the mean value with a standard deviation equal to the associated error. 
The uncertainty derived from the bootstrap method is 51\% for $E_G$, 39\% for $E_K$, and 59\% for $E_B$. 

The contributions of the three energy components are comparable, with the gravitational energy being the most significant, but not by a large margin.  
Given the field strength, the Alfv\'{e}n speed is, $v_A = B_\mathrm{tot} /\sqrt{4\pi \rho} = 2.3 \pm 0.8 \; \mathrm{km \, s^{-1}}$. 
The relative contribution of turbulence and magnetic fields can be assessed by comparing their energies, $\beta = E_K/E_B = 3(\delta v_\mathrm{los}/v_A)^2 = 0.65$, implying that the magnetic energy dominates over turbulence energy.  
Using the virial theorem, we solve for the virial mass (Eq.~\ref{eq:mvir}) and obtain $M_\mathrm{vir} = 45 \pm 16 \; M_\odot$.
Comparing the virial mass, $M_\mathrm{vir}$, to the total mass, $M$, we further assess the dynamic state of W3~IRS5 with the virial parameter, $\alpha_\mathrm{vir} = M_\mathrm{vir}/M = 0.8 \pm 0.4$.   
A small value of the virial parameter, $\alpha_\mathrm{vir} < 1$, indicates that the clump is dominated by gravitational energy and allows an ongoing collapse feeding mass to the central protocluster.  
Another test for gravitational instability is to compare the free-fall time, $t_\mathrm{ff}$, with the dynamical crossing time, $t_\mathrm{cross} = 2R/\sqrt{3} \, \delta v_\mathrm{los}$, which represents the timescale for a signal to transmit from one side of the core to the other side \citep{McKee2007}. 
%where the velocity dispersion of the gas represents the effective sound speed.  
A clump becomes gravitationally unstable if the free-fall time is less than its dynamical crossing time, $t_\mathrm{ff} < t_\mathrm{cross}$.
Given the clump radius of $R = 0.05 \; \mathrm{pc}$ and $\delta v_\mathrm{los} = 1.07 \; \mathrm{km \, s^{-1}}$, the dynamical crossing time for the clump is $t_\mathrm{cross} = 5.3 \times 10^4 \; \mathrm{yr}$.  
The free-fall time corresponds to the mean density of $n_\mathrm{H_2} = 10^6 \; \mathrm{cm^{-3}}$ is $t_\mathrm{ff} = \sqrt{3\pi/32 G \rho} = 3.4 \times 10^4 \; \mathrm{yr}$. 
Since $t_\mathrm{ff}/t_\mathrm{cross} = 0.6$, W3~IRS5 is likely collapsing.  

%%%
\subsection{Stellar Feedbacks in a Cluster Environment} % 
High-mass stars are born in cluster environments and always alter their surroundings through stellar feedback, such as ionization, radiative heating, jets/outflows, stellar winds, etc.  
In W3~Main, at least ten \HII regions are present at various evolutionary stages, from hypercompact, ultracompact, compact, to diffuse \HII regions \citep{Tieftrunk1997}. 
Based on X-ray {\it Chandra} observations, \citet{Feigelson2008} revealed a large cluster that extends over 7~pc with $\sim 900$ pre-main-sequence stars of typical age $> 0.5 \; \mathrm{Myr}$ in a nearly spherical distribution centered on the \HII regions. 
Using NIR photometric and spectroscopic data, \citet{Bik2012} studied the OB stars associated with these \HII regions and found the most massive star, IRS2 ($M_* = 35 M_\odot$), associated with W3~A at an age of 2$-$3~Myr.  
The latest star formation occurs in W3~IRS5, the only source detected with circumstellar emission in the $K$-band \citep{Bik2012}.  
In contrast, the lifetime for UC\HII regions, such as W3~F, is roughly $10^5 \; \mathrm{yr}$, and even shorter ($\sim 10^4 \; \mathrm{yr}$) for HC\HII regions embedded in W3~IRS5.   
%The lack of extinction in these OB stars suggests a quick dispersal of the circumstellar material within $10^5 \; \mathrm{yr}$. 
A plausible sequential star formation scenario has been proposed to describe the age spread of at least 2$-$3~Myr in the stellar population over a prolonged formation time in W3~Main \citep{Bik2012,Wang2012}

W3~IRS5 is surrounded by three \HII regions (Fig.~\ref{fig:bmod}), with the relatively diffuse W3~A C\HII region to the east and the W3~B C\HII region to the west.  
The molecular gas shows a ring-like structure around W3~A with a velocity gradient, suggesting that the \HII region expands and interacts with the ambient molecular gas \citep{Wang2012}.  
Likewise, the W3~B \HII region displays a bright limb on the southeastern edge and a velocity gradient in the H66$\alpha$ recombination line, indicating interaction with the ambient gas \citep{Tieftrunk1997}. 
The $\mathrm{^{13}CO}$ emission \citep{Wang2012} shows a ridge of molecular gas along the edge of W3~A  and parallel with the background $B_\mathrm{pos}$ at $\mathrm{P.A.} \sim 140^\circ$.  
In a cluster environment with high-mass stars, an expanding \HII region may drive ionization and shock fronts into adjacent molecular clouds and sweep up on its edge a layer of dense neutral material that eventually becomes gravitationally unstable \citep{Elmegreen1977}.   
W3~IRS5 was likely formed out of the dense gas ridge plowed by the expansion of W3~A, and perhaps also W3~B.          
Once the contraction started, gravity pulled the magnetic field lines inward, creating a pinched morphology.  
The subsequent dynamic interactions between the protostars may cause their outflow axes to become misaligned with the initial magnetic field \citep{Zhang2014}.   
Meanwhile, the O-type star, IRS7, ionized its surroundings and carved a photoionized arc into the denser neutral gas around W3~IRS5, reshaping the magnetic fields with concave distortion in the southern region.  
The cometary shape of W3~F with a bright limb on the northern edge also supports an interaction scenario of a bow shock created by IRS7 moving towards the dense gas surrounding W3~IRS5.
With the large mass reservoir of $715 \; M_\odot$ associated with W3~IRS5 \citep{Wang2012}, the collapsing core is still feeding the central protocluster, but the impact caused by the expansion of W3~F may induce more instabilities at a later time.

%%%%%%%%%%%%%
\section{Summary \label{sec:summary}} %
We have conducted polarimetric observations with the SMA at $340 \; \mathrm{GHz}$ with an angular resolution of $2\farcs7 \times 2\farcs2$ to map the polarized dust emission in the nearby ($1.83 \; \mathrm{kpc}$) luminous ($2 \times 10^5 \; L_\odot$) high-mass star-forming region W3~IRS5.  
The projected magnetic fields, $B_\mathrm{pos}$, appear organized in two parts: a pinched morphology centered at SMM2 in the northern part and a concave shape centered at IRS~7, the star powering the UC\HII region W3~F to the south.   
We fit the observed magnetic field morphology with a model composed of two components: an SFF model for the hourglass geometry in the northern part and a spherical model with uniform azimuthal components in the southern part.  
Using the DCF method, we calculated a projected field strength of $B_\mathrm{pos} = 1.36 \; \mathrm{mG}$.  
Combining with $B_\mathrm{los}$ from the Zeeman measurement, we estimated a magnetic field strength of $B_\mathrm{tot} = 1.6 \; \mathrm{mG}$.  

To assess the relative importance of gravity, turbulence, and magnetic fields, we compare their energy contributions and find that gravitational energy is the most dominant, followed by magnetic energy, and then turbulent energy.
We also test gravitational instability with the virial parameter $\alpha_\mathrm{vir}$ and the ratio of the free-fall time, $t_\mathrm{ff}$ to dynamical crossing time, $t_\mathrm{cross}$.  
The small values of $\alpha_\mathrm{vir} = 0.8$ and $t_\mathrm{ff}/t_\mathrm{corss} = 0.6$ suggest an ongoing collapse in W3~IRS5.
The morphology of magnetic fields and the proximity of the expanding \HII regions put forward a scenario for W3~IRS5. 
A dense core formed within a neutral gas ridge plowed by the expansion of W3~A, and perhaps also W3~B.  
The core became gravitationally unstable and started the contraction, where gravity pulled the magnetic field lines inward, resulting in the pinched field morphology.  
Subsequent expansion of W3~F, ionized by the O-type star, IRS7, perturbed the magnetic field with concave patterns in the southern part. 
Due to the high stellar density in W3~IRS5, dynamical interactions among protostars cause outflows to be misaligned.  

We also mapped molecular outflows in the $\mathrm{CO} \; (3-2)$ and $\mathrm{SiO} \; (8-7)$ emissions.  
Except outflow-a associated with the large-scale CO bipolar outflow \citep{Mitchell1991}, three well collimated outflows are all detected as previously reported in the $\mathrm{CO} \; (2-1)$ emission \citep{Wang2012}. 
The outflows appear more collimated and have more extensive velocity spreads in our observations due to higher excitation energies.  
The most prominent bipolar outflow is outflow-b along east-west.    
A newly discovered compact knot at the farthest position of the red-shifted flow has a velocity offset reaching up to $\Delta v = 100 \; \mathrm{km \, s^{-1}}$, giving a dynamical timescale of 820~yr.    
The outflow-d shows an asymmetric structure with a more dominant red-shifted flow.  
The bipolar outflow along the line of sight, outflow-c, is also detected in our data with an extensive velocity range from $\Delta v = -63$ to $80 \; \mathrm{km \, s^{-1}}$.  

We found no rotation at the clump scale, but tentative velocity gradients in two distinct directions within $1^{\prime\prime}$ near SMM2 were discerned by finding peak positions channel by channel with 2D Gaussian fits.  
Similar to previous studies by \citep{Wang2013}, the peak position moves from south to north at $\mathrm{P.A.} = 30^\circ$, then from west to east at $\mathrm{P.A.} = 122^\circ$ with increasing velocity.  
These velocity gradients are often related to unresolved kinematics of outflows or rotating disks.  
Observations with higher angular resolutions are required to reveal the nature of these velocity gradients.

%% Please use the acknowledgment and contribution environments. This will 
%% be anonomyized when the "anonymous" style option is used. 
\begin{acknowledgments} %
We thank the anonymous referee for the constructive comments that help improve the manuscript. 
H.V.C. acknowledges financial support from National Science and Technology Council (NSTC) of Taiwan grants NSTC 112-2112-M-007-041.
The Submillimeter Array is a joint project between the Smithsonian Astrophysical Observatory and the Academia Sinica Institute of Astronomy and Astrophysics and is funded by the Smithsonian Institution and the Academia Sinica. We recognize that Maunakea is a culturally important site for the indigenous Hawaiian people; we are privileged to study the cosmos from its summit.  
\end{acknowledgments} %

%\begin{contribution} %
%\end{contribution} %


\vspace{5mm} %
\facilities{SMA} %

%% authors a place to specify which programs were used during the creatjkion of 

\software{% 
%astropy \citep{AstropyCollaboration2013,AstropyCollaboration2018},  
%MIR \citep{}, 
MIRIAD \citep{Sault1995}
} %



\appendix %
%%%%%%%%%%
\section{The observed and synthetic Stokes $I$, $Q$, and $U$ maps \label{sec:iqumap}} %
\begin{figure}[h!] %
\epsscale{1.0} %
\plotone{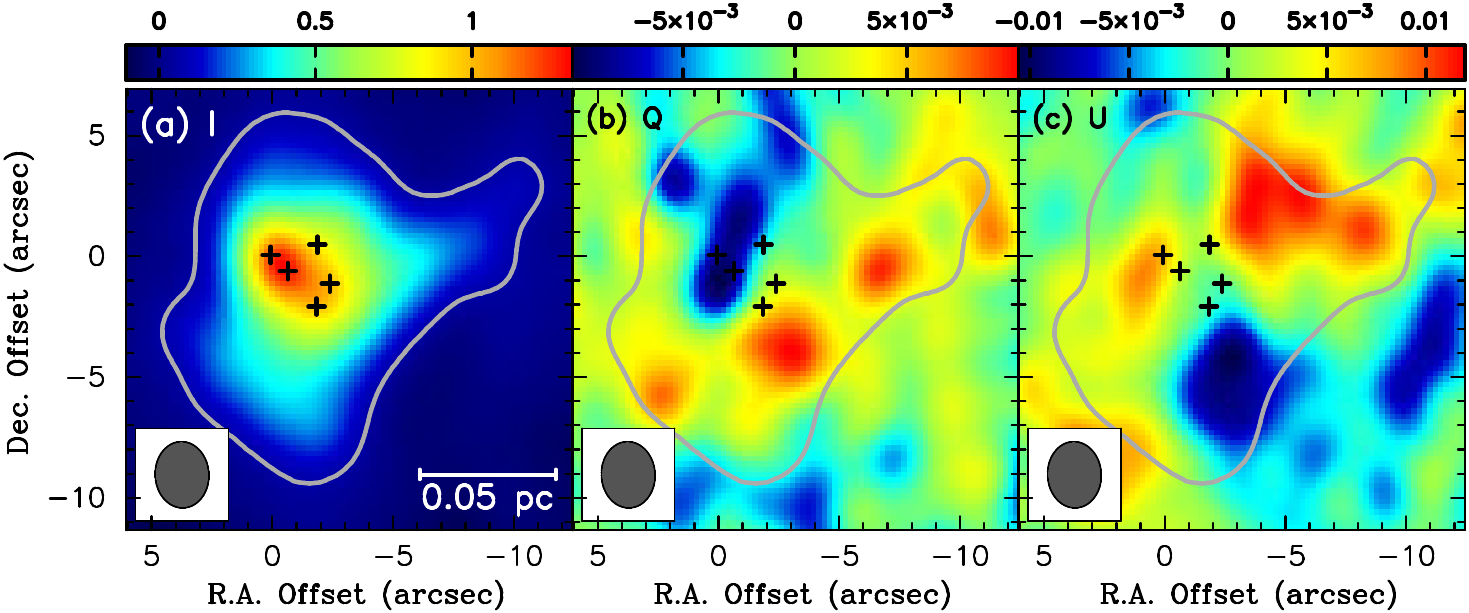} %
\caption{ 
Polarization maps of the (a) Stokes-$I$, (b) Stokes-$Q$, and (c) Stokes-$U$ parameters.   
The thick gray contour outlines the $340 \; \mathrm{GHz}$ continuum emission at the $3\sigma$ level. 
\label{fig:iqu}} %
\end{figure}

Figure~\ref{fig:iqu} presents the observed Stokes $I$, $Q$, and $U$ maps. 
%% Fig. 5

\begin{figure}[h!] %
\epsscale{1.0} %
\plotone{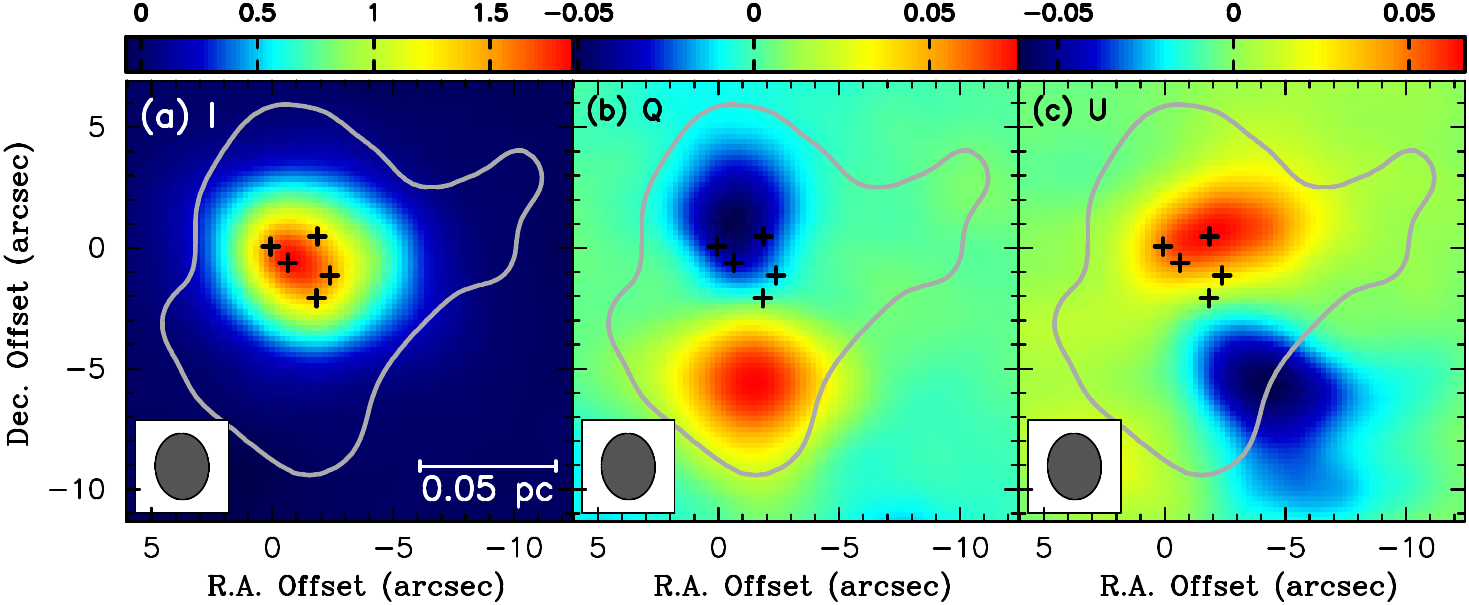} %
\caption{ 
Synthetic polarization maps of the optimized model for the (a) Stokes-$I$, (b) Stokes-$Q$, and (c) Stokes-$U$ parameters.   
The thick gray contour outlines the $340 \; \mathrm{GHz}$ continuum emission at the $3\sigma$ level. 
\label{fig:modiqu}} %
\end{figure}

Figure~\ref{fig:modiqu} presents the synthetic Stokes $I$, $Q$, and $U$ maps of the optimized model. 
The comparison between the observed data and the synthetic results uses the position angles of the magnetic field orientations, $\psi_\mathrm{obs}$ and $\psi_\mathrm{mod}$, which are related to the ratio $Q/U$. 
The intensities of the Stokes-$I$, $Q$, and $U$ images are not constrained during the optimization process. 
The polarized intensities, Stokes-$Q$ and $U$, are mainly scaled with $\alpha$ (Eq.~\ref{eq:alpha}), which is usually assessed by multi-wavelength studies \citep[e.g.][]{Draine2021a,Draine2022}.    
We explore the dependence of the model Stokes-$Q$ and $U$ images with $\alpha$ and find the model images approach to the observed intensity levels when $\alpha \sim 0.03$, which is much smaller than the estimate value of $0.17$ obtained from the over-simplified approximation (Eq.~\ref{eq:alpha_pmax}).
However, we emphasize that our observation at a single frequency cannot constrain $\alpha$.  

%% Fig. 6

%%%%%%%%%%
\section{Velocity gradients in the $\mathrm{SO_2} \; 34_{3,31}-34_{2,32}$ emissions \label{sec:so2}} %
\begin{figure}[h!] %
\epsscale{1.0} %
\plotone{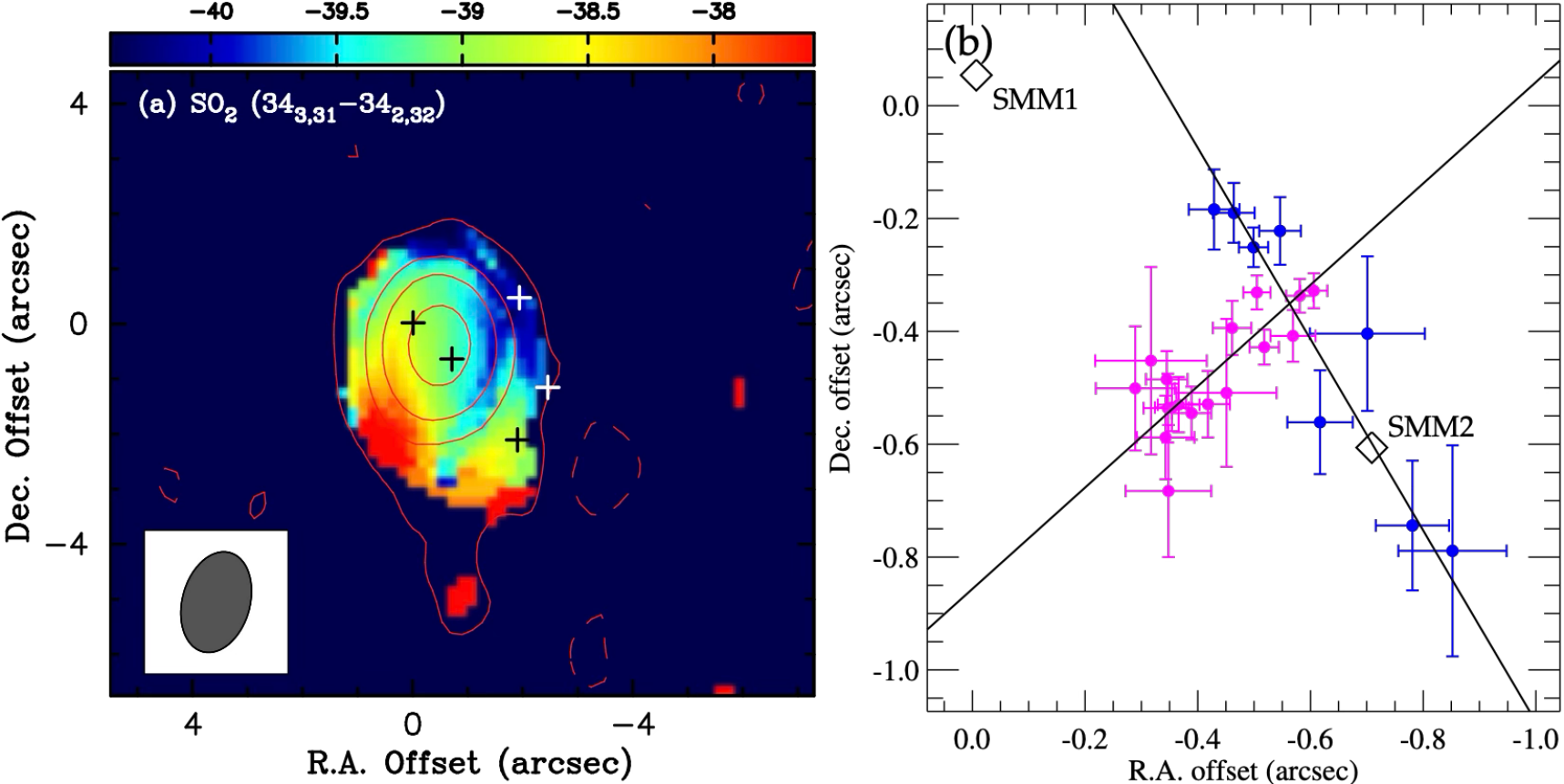} %
\caption{ 
(a) Integrated intensity map of the $\mathrm{SO_2} \; 34_{3,31} - 34_{2,32}$ emission (red contours) overlaid on the intensity-weighted velocity map (color scale).   
Red contours are plotted at $(-3,3,20,40,80) \times \sigma$, where $\sigma = 2.7 \; \mathrm{K \, km \, s^{-1}}$.
Crosses mark the peak positions of the five continuum sources \citep{Wang2013}.  
The beam is shown in the lower-left corner with a size of $1\farcs9 \times 1\farcs2$ (P.A. $=-16^\circ$). 
(b) Peak positions of the 2D Gaussian fits for channels with peak emission above $6\sigma$, where $\sigma = 0.23 \; \mathrm{Jy \, beam^{-1}}$.  
Black lines represent the linear regressions to peak positions with uncertainties.  
The derived P.A. $= 31^\circ$ and $132^\circ$ for data in the velocity range of $-45.0$ to $-42.0 \; \mathrm{km \, s^{-1}}$ (blue) and $-41.5$ to $-33.0 \; \mathrm{km \, s^{-1}}$ (red), respectively.  
The velocity increases from south to north (blue) and then from west to east (red).       
\label{fig:so2}} %
\end{figure}

We reanalysized the velocity gradients in two distinct directions reported by \citet{Wang2013}, particularly for the $\mathrm{SO_2} \; 34_{3,31}-34_{2,32}$ emission ($E_\mathrm{up} = 582 \; \mathrm{K}$).  
We were able to improve the image quality by including the visibilities in the extended configuration to achieve a sensitivity of $\sigma = 0.25 \; \mathrm{Jy \, beam^{-1}}$ with a smaller beam size of $1\farcs9 \times 1\farcs2$ 
and a better velocity resolution of $0.5 \; \mathrm{km \, s^{-1}}$.  
The intensity-weighted velocity map shows a complicated velocity distribution (Fig.~\ref{fig:so2}a).
We then perform similar analyses as described in Sect.~\ref{sec:vgrad} to identify any ordered motion in peak positions (Fig.~\ref{fig:so2}b). 
In addition to the previously reported velocity gradient along a line at $\mathrm{P.A.} = 132^\circ$ (red dots), we find the peak position in the lower velocity channels moving along the direction at $\mathrm{P.A.} = 31^\circ$ (blue dots), similar to the motion previously found in $\mathrm{HCN \; (4-3)} \, v_2 =1$.

We also insepct the $\mathrm{HCN \; (4-3)} \, v_2 =1$ ($E_\mathrm{up} = 1067 \; \mathrm{K}$) data but do not gain much improvement.  

%% Fig. 7

%%%%%%%%%%
\section{ Calculating the Stokes $I$, $Q$, and $U$ images \label{sec:iqueqn}} %
The polarization arising from the thermal emission of aligned dust grains is a unique tool to probe the projected component of magnetic fields.  
There is a myriad of literature on calculating polarization produced by spinning dust grains in the presence of magnetic fields under various assumptions for a range of astrophysical problems \citep[e.g.][]{Martin1974,Mishchenko1991,Whitney2002,Reissl2016,Lee1985,Wardle1990,Fiege2000a,Padoan2001,Goncalves2005,Frau2011,Padovani2012,Reissl2014,Reissl2016}.    
However, the notation and sign conventions differ among various authors.  
In this study, we basically follow the method and notation of \citet{Draine2022}.  
For clarity, we briefly outline the equations used to calculate the Stokes $I$, $Q$, and $U$ images. 

Interstellar dust grains spin rapidly around their short axes, i.e. the symmetry axis of the largest moment of inertia, and the spin axes tend to precess around local magnetic fields, $\vec{B}$ \citep{Lee1985,Draine2021a}.  
%For a grain ensemble, the  can be calculated as the averages over grain size and orientation distributions multiplied by the number density of dust grains, $n_d$.  
The polarization effects caused by a grain ensemble are represented by an elongated mean profile projected on the plane of the sky, which can be calculated as the averages over grain size and orientation distributions multiplied by the number density of dust grains, $n_d(s)$. 
The polarization of aligned dust grains can be described by their effects on the Stokes parameters\footnote{We follow the IAU convention for the Stokes parameters: positive $Q$ for the electric vector along the N-S orientation ($\mathrm{P.A.} = 0^\circ$); positive $U$ for the NE-SW orientation ($\mathrm{P.A.} = 45^\circ$); positive $V$ for the electric vector rotating counterclockwise (increasing in P.A.) toward the source.  }, $\vec{S} = [I,Q,U,V]^{T}$.  
Following \citet{Martin1974}, we first describe the polarization in a reference coordinate system fixed to the mean grain profile.  
When scattering is neglected, the radiative transfer equation including thermal emission of spheroid grains can be written in the Stokes vector form  \citep{Martin1974,Wardle1990,Whitney2002,Reissl2016,Draine2022}
\begin{equation} %
\frac{\dif}{\dif s} \mqty( I \\ Q \\ U \\ V ) = 
             \mqty( -\delta & \Delta \sigma & 0 & 0 \\
                         \Delta \sigma & -\delta & 0 & 0 \\
                         0 & 0 & -\delta & -\Delta \epsilon \\
                         0 & 0 & \Delta \epsilon & -\delta ) 
              \mqty( I \\ Q \\ U \\ V ) + B_\nu(T_d) \mqty(\delta \\ -\Delta \sigma \\ 0 \\ 0 )
           = \mqty( -\delta & \Delta \sigma & 0 & 0 \\
                         \Delta \sigma & -\delta & 0 & 0 \\
                         0 & 0 & -\delta & -\Delta \epsilon \\
                         0 & 0 & \Delta \epsilon & -\delta )  
               \mqty( I - B_\nu(T_d) \\ Q \\ U \\ V ),                 
\end{equation} %
We then consider grains aligned with a local magnetic field in the direction, $\vec{\hat{b}} \equiv \vec{B}/|\vec{B}|$, which is at an inclination angle $i_b$ relative to the line of sight, with the projected component oriented at a position angle, $\psi_b$.  
The transformation of the Stokes parameters to the general sky coordinate system is \citep{Martin1974,Draine2022}
\begin{equation} %
\frac{\dif}{\dif s} \mqty( I \\ Q \\ U \\ V ) = 
               \mqty( -\delta & \Delta \sigma \cos 2\psi_b & \Delta \sigma \sin 2\psi_b & 0 \\
                            \Delta \sigma \cos 2\psi_b & -\delta & 0 & \Delta \epsilon \sin 2\psi_b \\
                            \Delta \sigma \sin 2\psi_b & 0 & -\delta & -\Delta \epsilon \cos 2\psi_b \\
                            0 & -\Delta \epsilon \sin 2\psi_b & \Delta \epsilon \cos 2\psi_b & -\delta )
                 \mqty( I - B_\nu(T_d) \\ Q \\ U \\ V ), 
\label{eq:stokes} %              
\end{equation} %
where $\delta$, $\Delta \sigma$, and $\Delta \epsilon$ are used to describe the extinction, linear dichroism, and linear birefringence, respectively.  
Using the modified picket fence approximation, \citet{Draine2021a} relates $\delta$, $\Delta \sigma$, and $\Delta \epsilon$ to the alignment factor, $f_\mathrm{align}$, and the inclination angle, $i_b$: 
\begin{eqnarray} %
\delta &=& n_d \left[ C_\mathrm{ran} + f_\mathrm{align} C_\mathrm{pol} \left( \cos^2 i_b -\frac{1}{3} \right) \right] \\
\Delta \sigma &=& n_d f_\mathrm{align} C_\mathrm{pol} \sin^2 i_b \\
\Delta \epsilon &=& n_d f_\mathrm{align} C_\mathrm{cir} \sin^2 i_b , 
\end{eqnarray}
where $C_\mathrm{ran}$ is the absorption cross section for an ensemble of randomly oriented grains, $C_\mathrm{pol}$ and $C_\mathrm{cir}$ are the respective cross sections for the linear and circular polarization of the mean grain profile.  
Assuming that circular polarization is negligible, we drop all the quantities related to the Stokes-$V$ and linear birefringence, $\Delta \epsilon$.    
If we further assume that the grain properties are uniform, the equations for extinction and linear polarization can be simplified as  
\begin{eqnarray} %
\delta &=& n_d C_\mathrm{ran} \left[ 1 + \alpha \left( \cos^2 i_b -\frac{1}{3} \right) \right] \\
\Delta \sigma &=& n_d C_\mathrm{ran} \, \alpha \sin^2 i_b, 
\end{eqnarray}
where $\alpha$ is the ratio of the cross section for an aligned grain ensemble to that of a randomly oriented grain ensemble \citep{Fiege2000a,Padoan2001}:  
\begin{equation} %
\alpha = \frac{f_\mathrm{align} C_\mathrm{pol}}{C_\mathrm{ran}}.  \label{eq:alpha} %
\end{equation} %
The absorption cross section of a grain ensemble may be related to the dust opacity by $n_d C_\mathrm{ran} = \rho \kappa_\nu$.  
Hence, the matrix multiplication of Eq.~(\ref{eq:stokes}) leads to Eq.~(\ref{eq:rti})--(\ref{eq:rtu}).

The cross-section ratio, $\alpha$, represents the relative importance of linear polarization to absorption.  
Assuming a constant $\alpha$, we may estimate its value from the maximum degree of polarization, $p_\mathrm{max}$, in the data.  
Using Eq.~(62) of \citet{Draine2021a}, the maximum degree of polarization occurs when grains have a perfect alignment, $f_\mathrm{align}=1$, and  $\vec{\hat{b}}$ lies in the plane of the sky, $\sin^2 i_b = 1$.  
Then we have  
\begin{equation} %
p_\mathrm{max} = \frac{\alpha}{1 - \alpha/3}.   
\end{equation} %
or alternatively, 
\begin{equation} %
\alpha = \frac{p_\mathrm{max}}{1 + p_\mathrm{max}/3}  \label{eq:alpha_pmax}
\end{equation} %

%%%%%%%%%%
\section{Virial mass for a power-law density distribution with a central star \label{sec:mvir}} %
Consider a spherical clump of radius $R$ with a power-law gas density distribution, $\rho = \rho_0 r^{-p}$, that harbors a star cluster of mass $M_*$ at its center.    
Let the total mass of the clump be $M = M_* + M_g$, where $M_g$ is the gas mass in the clump, and denote the stellar fractional mass as $s \equiv M_*/M$.    
As the stellar mass grows in time, $s$ increases from $0$ to $1$.  
The enclosed mass within a radius of $r$ is given by
\begin{equation} %
m(r) = \int_0^{R} \rho(r^\prime) 4\pi r^{\prime 2} \dif r^\prime + M_* = M_g \left( \frac{r}{R} \right)^{3-p} + M_*,
\end{equation} %
where $M_g = 4\pi \rho_0 R^{3-p}/(3-p)$.  
The gravitational energy, $E_G$, can be calculated by integrating the contribution from individual spherical shells 
\begin{eqnarray} %
E_G &=& \displaystyle -\int_0^R \frac{G m(r^\prime) \rho(r^\prime) \, 4\pi r^{\prime 2} }{r^\prime} \dif r^\prime \nonumber \\
      &=& -\int_0^R \frac{G M_g (3-p)}{R^{3-p}} r^{\prime (1-p)} \left[ M_g \left( \frac{r^\prime}{R} \right)^{3-p} + M_* \right] \dif r^\prime \nonumber \\
      &=& -\frac{G M_g}{R} (3-p) \left[ \frac{M_g}{5-2p} + \frac{M_*}{2-p} \right]   \\
      &=&  -\frac{G M^2}{R} (3-p) (1-s) \left[ \frac{1-s}{5-2p}  + \frac{s}{2-p} \right], \hspace{1cm} \text{if } p \not= 2, \frac{5} {2}, 3 \label{eq:EG}
\end{eqnarray} %
In most star-forming cores, one may expect $p \lesssim 2$.
The observed Gaussian line width, $\delta v_\mathrm{los}$, measures the one-dimensional velocity dispersion along the line of sight.  
Assuming an isotropic velocity dispersion, the kinetic energy can be written as  
\begin{eqnarray} %
E_K &=& N \frac{3}{2} \mu m_\mathrm{H} \delta v_\mathrm{los}^2 = \frac{3}{2} M_g \delta v_\mathrm{los}^2 \\
        &=& \frac{3}{2} M \delta v_\mathrm{los}^2 (1-s),   
        \label{eq:EK}
\end{eqnarray} %
where $N$ is the total number of gas particles, $\mu = 2.33$ the mean molecular weight, and $m_\mathrm{H}$ the hydrogen mass.  
In the presence of magnetic fields, the energy stored in the fields is 
\begin{eqnarray} %
E_B &=& \frac{B^2}{8 \pi} \frac{4\pi}{3} R^3 = \frac{1}{6} B^2 R^3 = \frac{1}{2} M_g v_A^2 \\
        &=& \frac{1}{2} M v_A^2 (1-s), 
        \label{eq:EB}
\end{eqnarray} %
where $v_A = B/\sqrt{4\pi \rho} = B \sqrt{R^3/3 M_g}$ is the Alfv\'{e}n speed.  
For a system in equilibrium, the virial theorem requires the acceleration of the moment of inertia to vanish,  
\begin{equation} %
0 = \frac{1}{2} \frac{\dif^2 I}{\dif t^2} = 2 E_K + E_G + E_B. 
\end{equation} %
Solving the above equation for the total mass, $M$, we obtain the virial mass, $M_\mathrm{vir}$, as
\begin{equation} %
	M_\mathrm{vir} = \frac{R}{G} \frac{6 \delta v_\mathrm{los}^2 + v_A^2}{2 (3-p)} \left[ \frac{1-s}{5-2p} + \frac{s}{2-p} \right]^{-1}. 
	\label{eq:mvir}
\end{equation} %
The dynamical state of a clump is often evaluated by comparing the virial mass, $M_\mathrm{vir}$, to the clump mass, $M$, that is, the virial parameter, $\alpha_\mathrm{vir}$, 
\begin{equation} %
	\alpha_\mathrm{vir} = \frac{M_\mathrm{vir}}{M} = \frac{R}{G M}  \frac{6 \delta v_\mathrm{los}^2 + v_A^2}{2 (3-p)} \left[ \frac{1-s}{5-2p} + \frac{s}{2-p} \right]^{-1},  
	\label{eq:alphavir}
\end{equation} %
where $\alpha_\mathrm{vir} < 1$ implies a system dominated by gravitational energy and will collapse.  
It is also useful to inspect how $M_\mathrm{vir}$ changes with the assumed density profile index, $p$, and the stellar mass fraction, $s$.  
When the velocity dispersion, $\delta v_\mathrm{los}$, and the Alfv\'{e}n speed, $v_A$, are fixed, the virial mass, $M_\mathrm{vir}$,  decreases with increasing $p$ and $s$, i.e. steeper density profile and larger stellar mass fraction, respectively. 
Yet the temperature and magnetic field strength of actual star-forming clumps increase with time, so the virial mass will not decrease monotonically.   

\bibliography{myrefs}{} %
\bibliographystyle{aasjournal} %



\end{document}